\documentclass[fleqn,usenatbib]{mnras}

\usepackage{newtxtext,newtxmath}

\usepackage[T1]{fontenc}

\DeclareRobustCommand{\VAN}[3]{#2}
\let\VANthebibliography\thebibliography
\def\thebibliography{\DeclareRobustCommand{\VAN}[3]{##3}\VANthebibliography}

\usepackage{graphicx}	
\usepackage{amsmath}	
\usepackage{siunitx}
\usepackage{longtable}
\usepackage{subcaption}
\usepackage[longtable]{multirow}
\usepackage{longtable}
\usepackage{booktabs}
\usepackage{tabularray}
\usepackage{cleveref}
\usepackage{pdflscape}
\usepackage{array}
\usepackage{ragged2e}

\title[C- and X-complex families]{Searching for primitive, dark, spectrally red asteroid families in the main belt with Gaia}

\author[U. Bhat et al.]{
Ullas Bhat$^{1}$\thanks{E-mail: uub1@leicester.ac.uk (UB)},
Chrysa Avdellidou$^{1}$,
Marco Delbo$^{2,1}$,
Thomas Dyer$^1$
\\
$^{1}$University of Leicester, School of Physics and Astronomy, University Road, LE1 7RH, Leicester, UK\\
$^{2}$Université C\^ote d’Azur, CNRS–Lagrange, Observatoire de la C\^ote d’Azur, CS 34229 – F 06304 NICE Cedex 4, France
}

\date{Accepted 19 November 2025. Received 16 September 2025}

\pubyear{\the\year{}} 

\begin{document}
\label{firstpage}
\pagerange{\pageref{firstpage}--\pageref{lastpage}}
\maketitle

\begin{abstract}
Dark asteroids with featureless neutral to red spectra are of particular interest due to their ability to potentially harbour primitive, hydrated, and possibly organic-rich material. These asteroids belong to the spectroscopic C-complex, to the X-types with low geometric visible albedo values as well as to the T- and D-type end members of the Bus-DeMeo spectroscopic taxonomy. Here we used Gaia Data Release 3 visible reflectance spectra to study the average spectral profiles of the C- and X-complex asteroid families in the central and outer main belt (orbital semi-major axis between 2.5 - 3.7~au). We found that eight of these families, namely 96 Aegle, 627 Charis, 1484 Postrema and 5438 Lorre, previously classified as C-complex families, and 322 Phaeo, 1303 Luthera, 5567 Durisen and 53546 2000BY6 previously classified as X-complex families, have redder slopes than implied by their previous classification and could be better classified as T-/D-type families. Some of these families may also feed the near-Earth asteroid population, being responsible for the observed T-/D-type excess. However, the analysis of their principal components of Gaia Data Release 3 spectra suggest that further near-infrared observations are needed in order to verify this identification.
\end{abstract}

\begin{keywords}
minor planets, asteroids: general -- methods: observational -- techniques: spectroscopic
\end{keywords}

\section{Introduction}
\label{intro}
The asteroid main belt, located between Mars and Jupiter, contains over a million known asteroids exhibiting a wide range of physical properties. In particular, asteroids show compositional diversity, suggesting that planetesimals, the first sizeable bodies that accreted in the beginning of the solar system formation and the building blocks of planets, formed at varying heliocentric distances and/or at different times. For instance, meteorite studies analysing fragments of these early planetesimals \citep[e.g,][]{kallemeyn1986, zhang1995} indicate that enstatite chondrites accreted in the terrestrial planet region. This is supported by the mainly enstatitic bulk compositions of both Mercury \citep[][and references therein]{ebel2011,nittler2018} and Earth \citep{javoy2010,warren2011,dauphas2017}. In contrast, ordinary chondrites are believed to have accreted just beyond the orbit of Mars \citep{warren2011,burkhardt2016}. Meanwhile, carbonaceous chondrites and other dark, primitive, ungrouped meteorites are thought to originate from greater heliocentric distances, beyond Jupiter's orbit \citep{warren2011,kruijer2017,nakamura2023}. It is assumed that Jupiter itself acted as a barrier in the early solar system, dividing the accretion regions of carbonaceous (CC) and non-carbonaceous (NC) chondritic materials \citep{kruijer2017}. Despite this separation, both reservoirs included differentiated bodies, as evidenced by the discovery of iron meteorites originating from both CC and NC sources \citep[e.g.][]{budde2016,kruijer2017}. The asteroid belt has been shown to be compositionally mixed with objects assumed to have come from both the inner and outer solar system \citep{demeo2014}. A compositional gradient is also known to exist with ordinary chondrite-like objects preferentially populating the inner portion of the main belt and CC-like objects predominantly in the outer regions \citep{gradie1982}. The presence of such a wide range of compositions within the relatively narrow region of the main belt can be explained by the implantation of small bodies through early dynamical processes in the solar system. This implantation can be driven by the giant planet instability \citep{tsiganis2005}, which is now recognised as a key mechanism responsible for transporting enstatite-rich material from the terrestrial region \citep{avdellidou2024}, as well as dark (low geometric visible albedo, $p_\text{V}$), primitive material from the trans-Neptunian region \citep{levison2009}. 

The vast majority of current main belt asteroids are collisional fragments of the residual planetesimal population—whether local or implanted \citep{delbo2017,dermott2018,delbo2019,ferrone2023}. Studying the physical properties of asteroid collisional family members provides valuable insights into the original composition and internal structure of planetesimals. When a planetesimal undergoes catastrophic disruption, the resulting fragments of the collisional family would sample materials from various layers of the original planetesimal. If the planetesimal was undifferentiated, we would expect all the resulting fragments to be homogeneous with chondritic composition. In contrast, a differentiated planetesimal, altered by internal heating from the decay of $^{26}$Al and $^{60}$Fe, would develop distinct layers, such as a metallic core, olivine-rich mantle, and a crust of chondritic or basaltic material depending on the degree of differentiation \citep{neumann2012,weiss2013}. The resulting asteroid family would then be compositionally heterogeneous, with fragments reflecting the composition of the layer from which they originated. While in the past most of the asteroid families were thought to be homogeneous, there is recent growing evidence of families created by the break-up of differentiated or partially differentiated bodies \citep[e.g.][]{oszkiewicz2015,galinier2024,avdellidou2025}. Such studies also enable us to establish connections with meteorite classes, estimate the parent bodies' original sizes, and infer the heliocentric distances at which they accreted \citep{trieloff2022}.

Based on the similarity between asteroid spectra, various taxonomic schemes have been proposed. The Tholen \citep{tholen1984} and Bus \citep{bus1999,bus2002a} taxonomies were developed on photometric and spectroscopic data in the visible (VIS) light range of the spectrum. 
\citet{demeo2009} extended the work of \citet{bus2002a} to the near-infrared (NIR) wavelengths, producing the widely used Bus-DeMeo (BDM) taxonomy. The BDM taxonomic scheme classifies the spectra in three main complexes (S, C, and X): The S-complex includes asteroids with geometric visible albedo ($p_\text{V}$) values $>0.12$ \citep{delbo2017} and spectra that show strong absorption features at $\approx$0.9~\SI{}{\micro\meter} and $\approx$2~\SI{}{\micro\meter}, the C-complex includes dark asteroids ($p_\text{V} < 0.12$) with spectra showing neutral slopes, while those asteroids belonging to the X-complex have redder slopes and a wide range of $p_\text{V}$. Both C- and X-complex asteroids have either small or no absorption bands. In addition to complexes there are nine other end-member types. Though the BDM taxonomic scheme is based only on the spectral characteristics of asteroids, the $p_\text{V}$ of asteroids have also been shown to be correlated with the taxonomic types, with $p_\text{V}$ of asteroids in the main belt having a bimodal distribution roughly splitting the S- and C-complexes at $p_\text{V}\mathbin{\approx}0.12$ \citep{usui2013,delbo2017}.

Spectroscopic and albedo analyses of both main belt asteroids and meteorites have allowed the scientific community to establish broad associations between ordinary chondrites and spectroscopic S-complex\footnote{In this work the preferred taxonomic scheme is the BDM taxonomy \citep{demeo2009}.} asteroids, while carbonaceous chondrites are linked to C-complex asteroids \citep[][and references therein]{reddy2015,demeo2022}. The X-complex asteroids encompass a variety of compositions, including enstatite chondrites \citep[linked to Xc-types,][]{vernazza2009,avdellidou2022}, enstatite achondrites \citep[linked to Xe-types,][]{gaffey1992}, iron meteorites \citep[linked to moderate albedo X-types,][]{cloutis1990}, mesosiderites \citep[linked to Xk-types,][]{vernazza2009}, and primitive carbonaceous materials \citep[linked to low-$p_\text{V}$ X-types or P-types in the Tholen taxonomy,][]{demeo2015}. Building on these broad associations, state-of-the-art research aims to link specific meteorite types to their source asteroid families \citep[see][for an overview]{jenniskens2025}. In some cases, these associations are unique, while in others, they remain ambiguous. Notable examples include the howardite-eucrite-diogenite (HED) meteorites with the Vesta family \citep[V-type,][]{russel2012}, aubrites with the Hungaria family \citep{lucas2019}, enstatite chondrites with the Athor family \citep{delbo2019,avdellidou2022}, iron meteorites and/or pallasites with the Kalliope family \citep{avdellidou2025}, L chondrites with the Massalia family \citep{marsset2024}, with the S-complex component of the Nysa-Polana, Juno and Gefion families being additional sources \citep{ciocco2025}, and LL chondrites with the Flora family \citep{vernazza2008}.

Asteroids belonging to the C-complex, the low-albedo X-type, and the T-/D-types of the end members are of particular interest due to their potential to carry primitive, hydrated, and possibly organic-rich material \citep{gradie1980,nationalresearchcouncil2007,glavin2015}. These bodies are part of the dark ($p_\text{V} < 0.12$), primitive small body population and are characterised by featureless spectra with increasing reflectance for increasing wavelengths, and a general absence of prominent absorption bands (or has very weak bands as is the case for Ch- and Cgh-types in VIS) in both the VIS and NIR wavelengths. Although a progressively increasing spectral slope is observed across the primitive, dark asteroid populations (X/P, T and D-types in increasing order of spectral slope), the underlying cause of this variation remains unclear. The D-type asteroids are thought to have formed beyond 15~au \citep{fujiya2019} and subsequently migrated inwards during the giant planet instability \citep{levison2009}. On the other hand, the origin of T-type asteroids and their connection to D-types, is still not clear. Is the spectral diversity indicative of compositional differences among these asteroid classes, or does it instead reflect the effects of space weathering processes \citep{fornasier2007}? With only a few links between known meteorites and D-type asteroids, such as the Tagish Lake \citep{brown2000, hiroi2001}, WIS 91600 \citep{hiroi2005}, MET 00432 \citep{nakamura2013}, and the Tarda \citep{marrocchi2021} meteorites, this issue is still unresolved. Previous spectroscopic studies of individual dark asteroids with red spectral slopes suggest that slope variability may result from the exposure of fresher surface material, possibly due to recent impacts, causing surfaces to appear spectrally redder \citep{fornasier2007,avdellidou2021b,hasegawa2022}.

Interestingly, a debiased overabundance of D-type asteroids has been observed in near-Earth space relative to the main belt \citep{perna2018,marsset2022}. 
Observational surveys indicate that these near-Earth D-type asteroids mainly originate from the 3:1J and 5:2J mean motion resonances (MMR) with Jupiter \citep{marsset2022}, suggesting the existence of a source population—presumably asteroid families—formed by the collisional disruption of D-type parent bodies near those resonances. However, previous spectroscopic and spectrophotometric surveys have failed to identify any such primitive, dark, red-sloped family within the main belt \citep{demeo2014}.

In this study, we primarily utilise asteroid VIS reflectance spectra of solar system objects from Gaia Data Release 3 (DR3) \citep{gaiacollaboration2023} to characterise all known C- and X-complex asteroid families in the central and outer main belt. We exclude the inner main belt C- and X-complex families, as they have already been comprehensively examined through ground-based observations \citep{deleon2016,pinilla_alonso2016,morate2016,morate2018,morate2019,arredondo2020,arredondo2021b,delbo2019,avdellidou2022} and Gaia DR3 data \citep{delbo2023}. Section \ref{sec:data} outlines the datasets used, including asteroid family memberships and Gaia DR3 spectra; Section \ref{sec:analysis} presents our analysis and results; and Section \ref{sec:discussion} discusses the implications of our findings.


\section{Data}
\label{sec:data}
We retrieved 55 asteroid families of the central and outer main belt from the catalogue presented in \citet{nesvorny2015} (hereafter NES15), which were thought to belong to the spectroscopic C- and X-complexes (\autoref{tab:family-list}). In the case of 7481 San Marcello family we used the updated definition and family membership by \citet{broz2022}, and hereafter will be referred as the 22 Kalliope family.
We also retrieved the family membership for the above C- and X-complex families from the Asteroids Family Portal\footnote{\url{http://asteroids.matf.bg.ac.rs/fam/}(accessed in October 2025)} \citep{novakovic2019} (hereafter AFP25). This membership was available for 41 of the 55 NES15 families. Moreover, we also retrieved the membership for three additional families that are not reported in NES15, namely 86 Semele, 727 Nipponia, and 1521 Seinajoki.
We obtained proper orbital elements ($a_\text{p}$, $e_\text{p}$, $i_\text{p}$), diameters, $p_\text{V}$ and spectral classes for family members from the Minor Planet Physical Properties Catalogue\footnote{mp3c.oca.eu} \citep[MP3C,][]{delbo2022}. The proper elements in MP3C are retrieved from \citet{knezevic2012}. The diameters (and the derived $p_\text{V}$) reported in MP3C are collected from various sources such as IRAS \citep{tedesco2002,ryan2010}, MSX \citep{ryan2010}, AKARI \citep{usui2011a}, WISE \citep{masiero2011}, NEOWISE \citep{mainzer2019}, stellar occultation campaigns \citep{durech2011}, and spatially resolved images using high-spatial resolution adaptive optics at large telescopes \citep{hanus2017}. For objects with multiple diameter and $p_\text{V}$ values reported in the literature, we used the average value weighted by the inverse of the square of the uncertainties.

We used Gaia DR3 \citep{gaiacollaboration2023} to retrieve VIS reflectance spectra for the members of the 54 out of the 55 NES15 families. There are no observations for the family members of 106302 2000UJ87. Gaia DR3 provides VIS reflectance spectra for 60,518 small bodies in the solar system in 16 discrete wavelength bands spanning 0.374-1.034~\SI{}{\micro\meter}, produced by averaging several epoch spectra \citep{gaiacollaboration2023}. Each epoch spectrum was produced by combining data from the blue and red spectrophotometers (BP and RP) on board the \textit{Gaia} spacecraft. The reflectance spectrum flag (RSF) available for each wavelength band and each spectrum provides information about the estimated quality of the band \citep{gaiacollaboration2023}. This quality flag can indicate if the data is validated (RSF=0), is suspected to be of poor quality (RSF=1), or is not good (RSF=2). The RSF was used to mask low-quality bands from the Gaia DR3 spectra for our analysis. Furthermore, we excluded the two reddest and two bluest data points from the analysis, as they have been shown to be affected by systematics due to the low efficiency of the spectrophotometers in these wavelength regions \citep{gaiacollaboration2023,galinier2023}.

For the subset of asteroids belonging to families and having Gaia DR3 spectrum, we also retrieved spectrophotometric data from the SDSS MOC4 dataset \citep{ivezic2002} when available. The SDSS dataset includes flux measurements in five spectral bands (u', g', r', i', z') centred at 0.354, 0.477, 0.623, 0.763, and 0.913~\SI{}{\micro\meter}, respectively \citep{carvano2010,gayon-markt2012}. These fluxes were converted to reflectance spectra following the procedure outlined in \citet{demeo2013}, excluding the u' band due to its high uncertainties.


\section{Analysis and Results}
\label{sec:analysis}

\subsection{Identifying interlopers}
\label{sec:interlopers}
The family memberships that we use throughout this study were determined using hierarchical clustering methods (HCM) in proper orbital element space. Thus, if an unrelated asteroid of a different spectral type than that of the bulk family happens to be in the same region of the proper orbital element space, it will be assigned as a member of the family. Since the C- and X-complex asteroids have mostly flat and featureless spectra, we manually inspected the reflectance spectra of all family members to identify possible S-complex interlopers, with large absorption features at ~0.9~\SI{}{\micro\meter}. By excluding the interlopers, we can obtain a cleaner average spectrum for each family which was then classified.

We also used the $p_\text{V}$ of the family members to aid in the identification of potential interlopers. For C-complex families, we expect most of the family members to have $p_\text{V} < 0.12$ \citep{delbo2017}. Thus, during manual inspection, if an asteroid had a spectrum with clear absorption bands and $p_\text{V} > 0.12$, we classified it as an interloper. However, if a spectrum did not have any clear absorption bands but had $p_\text{V} > 0.12$, we did not classify it as an interloper. We chose not to classify all asteroids with $p_\text{V} > 0.12$ as interlopers because X-complex families with metallic or enstatite compositions can also have $p_\text{V} > 0.12$ \citep{avdellidou2022,avdellidou2025}.



\subsection{Albedo distribution of families}
\label{sec:albedo}
The retrieved $p_\text{V}$ for the family members were used to investigate the $p_\text{V}$ distribution of each family \autoref{fig:all-albedo}. For most families, the $p_\text{V}$ distribution of the entire family and the distribution of the subset with available Gaia DR3 spectra are similar. We also observed that the outliers in these distributions were mostly the interlopers identified in Section \ref{sec:interlopers}, except for 145 Adeona. This family still contained some high $p_\text{V}$ interlopers after the manual inspection which is probably due to the contamination from the nearby S-complex Eunomia family. However, the number of these interlopers is small enough that they do not significantly affect the statistics calculated for the family. For families with more than \char`\~10 members with Gaia DR3 spectra, the $p_\text{V}$ distributions are approximately unimodal with a few exceptions that are detailed in Section \ref{sec:discussion}. The average $p_\text{V}$ of the entire families after removing interlopers and that of the subset with Gaia DR3 spectra for both NES15 and AFP25 catalogues are listed in \autoref{tab:family-list}.

\subsection{Average family spectra}
\label{sec:spectra}
After removing potential interlopers from each family, the spectrum of each asteroid is further cleaned by masking out all reflectances with RSF flags other than 0. Following the procedure described in \citet{delbo2023}, we then filtered for outliers by calculating the weighted median absolute deviation (WMAD) for each wavelength band per family using $1/\sigma^2$ as weights, where $\sigma$ is the reported uncertainty for each reflectance in Gaia DR3. All reflectance values with distances greater than $2.5\times$WMAD from the weighted median are masked out for further analysis. Using the remaining reflectance values, the average spectrum of the family was calculated by taking the weighted mean of the reflectance values in each wavelength band, using $1/\sigma^2$ as weights. The uncertainty for each wavelength band was calculated using a bootstrap method, where we randomly sampled 75\% of the reflectance values with replacement at each wavelength band and calculated the weighted mean. This procedure was repeated 1000 times, and the standard deviation of the means was used as the uncertainty for each wavelength band. The Gaia DR3 reflectance spectra are known to exhibit artificial reddening at smaller wavelengths, which was corrected by multiplying the mean reflectance spectra by a factor of 1.07, 1.05, 1.02 and 1.01 at wavelengths of 0.374, 0.418, 0.462 and 0.506~\SI{}{\micro\meter} respectively \citep{tinaut-ruano2023}. The calculated average reflectance spectra for all families for both catalogues are shown in \autoref{fig:all-spectra-nes} and \autoref{fig:all-spectra-nov}.

\begin{figure}
	\includegraphics[width=\columnwidth]{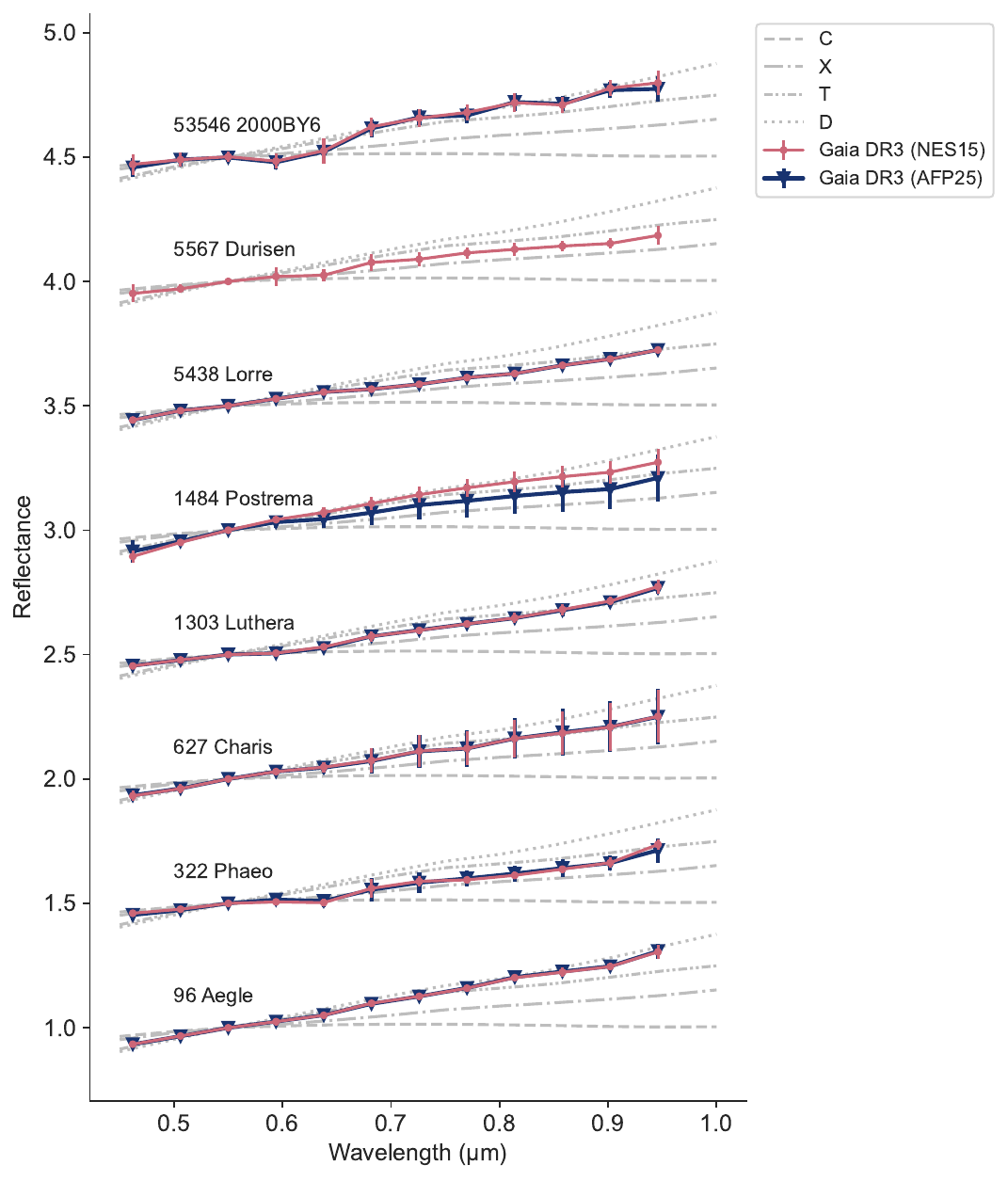}
	\caption{Average reflectance spectra of the reddest asteroid families using Gaia DR3. The spectra are normalised at 0.55~\SI{}{\micro\meter} and shifted on the y-axis by 0.5 between families for clarity. Where available, the average SDSS reflectance spectra are over-plotted.}
	\label{fig:part-average-spectra}
\end{figure}

Next we classified the average family reflectance spectra in the BDM taxonomic scheme \citep{demeo2009} by calculating the $\chi2$ value between the average family reflectance spectra and the BDM class templates \citep[see Appendix B.2 in][]{avdellidou2022}, and we report the first- and second-best classes as those with the lowest and second-lowest $\chi2$ values respectively in \autoref{tab:family-list}.
Our results show that from NES15, 12 families are classified within the C-complex and 27 within the X-complex for both first and second-best class. Five families are classified as C-complex in first-best class and X-complex in second-best class, and vice versa for another family. One family is classified as X-complex in the first-best class and T-type in the second-best class.
Similarly, from the AFP25, 10 and 20 families are classified within the C- and X-complexes respectively for both first and second-best class. Three families are classified as C-complex in the first-best class and X-complex in the second-best class. One family is classified as X-complex in the first-best class and T-type in the second-best class. There are eight families from NES15 that have T-/D-type as their first-best class, with five of these families also having T-/D-type as the second-best matching class. These eight families are 96 Aegle, 322 Phaeo, 627 Charis, 1303 Luthera, 1484 Postrema, 5438 Lorre, 5567 Durisen and 53546 2000BY6 (\autoref{fig:part-average-spectra}). Seven out of the eight families also have membership defined in AFP25, except 5567 Durisen, and are also classified as T-/D-type in their first-best class, with four of these families also having T-/D-type as the second-best matching class. Three additional families in AFP25 have also T-type as their first-best class, but we do not include them in our eight reddest families, as will be discussed further in Section \ref{sec:discussion}.

It has been shown that in several cases Gaia DR3 reflectance spectra are systematically redder than the corresponding spectra from Phase II of the Small Main-Belt Asteroid Spectroscopic Survey (SMASS). Specifically, the ratio of reflectances at $\lambda = 0.8932$~\SI{}{\micro\meter} and $\lambda = 0.748$~\SI{}{\micro\meter} is larger in Gaia DR3 than in SMASS \citep[][see in particular the right panel of their Figure 10, see also \citealt{galinier2023, galinier2024}]{gaiacollaboration2023}. This reddening effect could, for example, cause a T-type asteroid to appear as a D-type in Gaia DR3. According to \citet{gaiacollaboration2023}, this trend is unlikely to be due to phase-angle effects or to the choice of solar analogues in the DR3 processing, but may instead be related to calibration issues. Consistently, \citet{perna2018} investigated the dependence of spectra on phase angle for near-Earth asteroids and found that low-albedo, featureless objects, such as C- and D-types, do not exhibit detectable phase reddening.

Nevertheless, comparisons between Gaia DR3 and SDSS family average reflectances show good agreement in the spectral slopes at the red end for several families (e.g., 322 Phaeo, 1303 Luthera, 5567 Durisen, 53340 2000BY6 for the reddest, and others such as 22 Kalliope and 24 Themis). In some cases Gaia DR3 spectra are even bluer than SDSS, as for 87 Sylvia, 137 Meliboea, 627 Charis, and 3815 Konig in NES15. 

\subsection{Principal Component Analysis}
\label{sec:pca}
To minimize systematic differences between Gaia DR3 and other surveys, it is preferable to work solely within Gaia DR3 data, and to identify where the known spectral classes are located in the space of Gaia DR3 spectra. Principal Component Analysis (PCA) is a statistical technique used to reduce the dimensionality of a dataset while preserving as much information as possible. A robust approach is to reduce the dimensionality of Gaia DR3 reflectances using PCA, where the Gaia DR3 spectra can be transformed into a new set of orthogonal variables called principal components (PCs), with the first PC capturing the maximum variance in the data, followed by the second PC, and so on. Thereafter, we can map regions corresponding to the C-complex, X-complex and D-type asteroids in the (PC1, PC2) plane.

\begin{figure}
	\centering
	\includegraphics[width=\columnwidth]{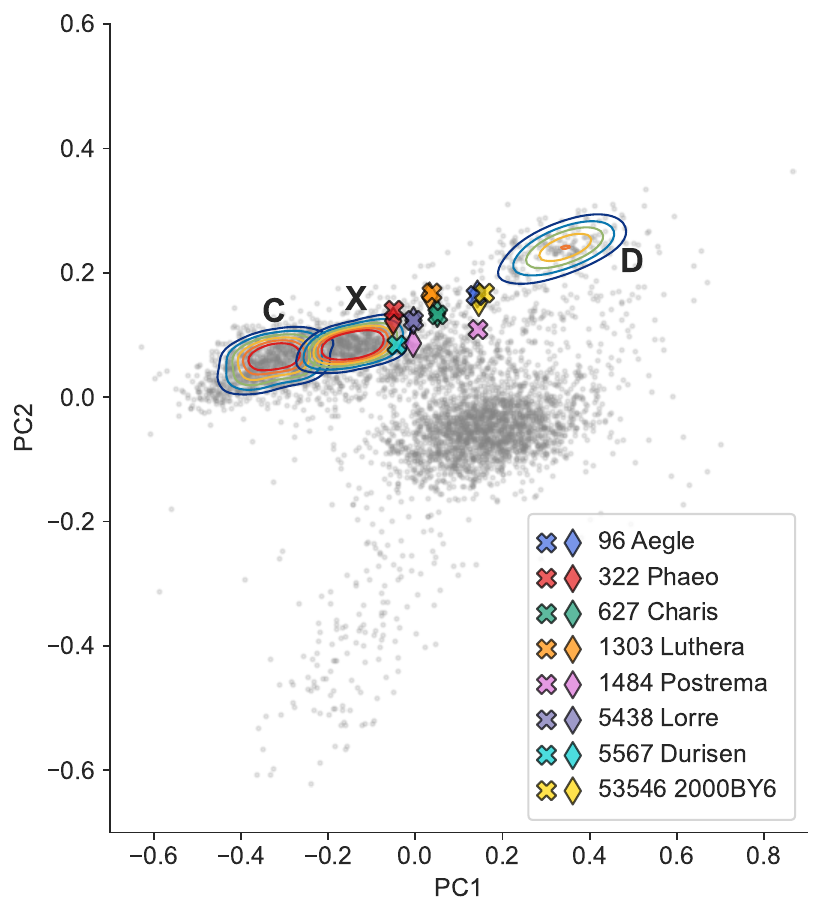}
	\caption{Location of the eight reddest families in the PC1-PC2 space in relation to the location of the C-complex, X-complex and D-type asteroids (depicted using KDE contours) with crosses and diamonds corresponding to NES15 and AFP25 catalogues respectively.}
	\label{fig:pca-candidates}
\end{figure}

To apply PCA to Gaia DR3 dataset, we followed similar procedure as in \citep{delbo2025}, however using only spectra with $\mathrm{SNR}>75$ \citep[signal-to-noise ratio as defined in][]{gaiacollaboration2023} to limit the impact of noise on the analysis.
The resulting distribution of first and second principal components (hereafter referred to as PC1 and PC2 respectively) is shown in \autoref{fig:pca-candidates}. Our PC1 roughly corresponds to the spectral slope and our PC2 captures information about the band depth at $\approx$0.9~\SI{}{\micro\meter} \citep[see Figure 9 of][for similarity]{gaiacollaboration2023}. The PC1-PC2 distribution shows various clusters which are known to correspond to different spectroscopic classes and complexes.


To investigate the relation between the clusters in PC1-PC2 space and spectroscopic complexes/classes we genarated three datasets of asteroids: one containing asteroids in Gaia DR3 with $\mathrm{SNR}>75$ that have previously been classified as C-complex from VIS and/or NIR spectroscopy, one for X-complex asteroids, and one for D-type asteroids (using spectral classes retrieved from MP3C), with 1340, 888 and 389 asteroids respectively. Though would be preferable to also have a dataset with the T-type asteroids, the number of known T-types in the literature with Gaia DR3 spectra is very small (only 30 asteroids with $\mathrm{SNR}>75$), thus we have excluded them from calculating Kernel Density Estimates (KDEs). However, since the T-type asteroids have slopes between those of the X-complex and D-types, we expect them to populate the region between the two in the PC1-PC2 space.
Using the \textit{scipy} Python package \citep{virtanen2020}, we calculated KDEs to give us the probability density of the distributions of the previously defined three datasets in the PC1-PC2 space (\autoref{fig:gaia-pca}), producing \textit{KDE}$_C$, \textit{KDE}$_X$ and \textit{KDE}$_{D}$, the KDEs for the C-complex, X-complex and D-type asteroids, respectively. The location of the maximum KDE probability density is different for the three datasets as expected. The \textit{KDE}$_C$ and \textit{KDE}$_X$ peaks are within the cluster of points centred at $\mathrm{PC1}\mathbin{\approx}-0.3$ and $\mathrm{PC2}\mathbin{\approx}0.1$, with their high probability density regions overlapping significantly. \textit{KDE}$_X$ peaks at higher values of PC1 as they have higher slope. The \textit{KDE}$_{D}$ peak corresponds to a clearly different cluster centred at $\mathrm{PC1}\mathbin{\approx}0.3$ and $\mathrm{PC2}\mathbin{\approx}0.25$.

When examining the location of the average spectra of the eight reddest families (96 Aegle, 322 Phaeo, 627 Charis, 1303 Luthera, 1484 Postrema, 5438 Lorre, 5567 Durisen and 53546 2000BY6) whose Gaia DR3 average reflectance spectrum was found to match with T- and/or D-types of the BDM, and compare it to the clusters in PC1-PC2 space corresponding to the C- and X-complexes and D-type asteroids, we found that they are all outside the C-/X-complex clusters, with Phaeo and Durisen located at the edge of the X-complex cluster, and Aegle and 2000BY6 located closest to the D-type cluster. The rest of the families are located where we expect the T-type cluster to be. Only Postrema has a significant difference in the average spectra in NES15 and AFP25 catalogues (\autoref{fig:part-average-spectra}), therefore its location in the PC1-PC2 space is also different between the two catalogues. The average reflectance spectrum for Postrema in AFP25 exhibits a shallower slope, placing its location further from the D-type cluster in the PC1-PC2 space. Nevertheless, it remains outside the X-complex cluster. This suggests that these families are spectrally distinct from the C- and X-complex asteroids, and may be more closely related to the T-/D-type asteroids. Their location also corresponds well with their classification presented in \autoref{tab:family-list}. However, none of these families have a centre in the region of the PC1-PC2 space where the known D-types from VIS and/or NIR spectroscopy are centred.

\section{Discussion}
\label{sec:discussion}
\subsection{Asteroid family average reflectance spectra}
\subsubsection{C-complex families}
The C-complex families according to this work are 10 Hygiea, 24 Themis, 31 Euphrosyne, 86 Semele, 128 Nemesis, 137 Meliboea, 145 Adeona, 656 Beagle, 668 Dora, 778 Theobalda, 1726 Hoffmeister, and 3438 Inarradas. Several of these families are spectroscopically well studied in the literature and were also classified in the C-complex \citep[i.e. Hygiea, Themis, Euphrosyne, Beagle, Hoffmeister, see][]{migliorini1996,fornasier2016,marsset2016,yang2020,depra2020b}.
Hoffmeister appears redder in the AFP25 catalogue compared to NES15. The Hoffmeister family has mostly noisy spectra in Gaia DR3 with only a few good quality spectra. In AFP25, there is one member with a good quality spectrum that appears much redder than the rest of the family members, increasing the slope of the weighted average to appear to match better with T-types, but also contributing to larger uncertainties. The SDSS average spectrum also agrees well with the average reflectance spectra of NES15. Thus, even though the average spectrum appears redder in AFP25, we classify it in the C-complex. Similarly, the Elfriede family, with only two members with Gaia DR3 spectra in both catalogues, appears redder in AFP25 due to one member with a redder spectrum. AFP25 also has one member with a SDSS spectrum which is the reddest compared to the Gaia DR3 spectra of the members. Since the family average reflectance spectra has smaller uncertainties in NES15, we consider this family to be in the C-complex.
This C-complex classification is in agreement with their dark $p_\text{V}$ distribution (\autoref{fig:all-albedo}). However, Nemesis, Hoffmeister and Adeona show contamination from higher albedo asteroids, which in the latter case most probably originate from the nearby S-complex Eunomia family.

There are five families that are classified as C-complex as the first-best class and X-complex as second-best class using the NES15 catalogue; these are 490 Veritas, 845 Naema, 1668 Hanna, 3815 Konig, and 4203 Brucato.
Naema is classified as C-complex in both first- and second-best class using AFP25.
Visual inspection of the average reflectance spectra of these families corroborates this classification. The average reflectance spectra of these families in Gaia DR3 and SDSS are also in good agreement in most cases.
In the case of the Konig family, the SDSS average reflectance spectrum is redder and is similar to that of X-types. The family average reflectance spectrum using AFP25 is redder than the one that results using the NES15, though not as red as the SDSS average reflectance spectrum. In contrast, the Brucato family appears bluer using AFP25 compared to NES15, with the SDSS average reflectance spectrum agreeing well with the NES15 Gaia DR3 average reflectance spectrum.
The $p_\text{V}$ distributions of these five families are all dark ($<0.12$) and Konig appears to be the darkest of all. Further observations are needed to disentangle the nature of Konig and determine if it is a C-complex family or a dark X-type one (or P-type in the Tholen taxonomy).

\subsubsection{X-complex families}
The X-complex according to this work families are 22 Kalliope, 81 Terpsichore, 87 Sylvia, 144 Vibilia, 283 Emma, 293 Brasilia, 363 Padua, 369 Aeria, 375 Ursula, 396 Aeolia, 410 Chloris, 569 Misa, 589 Croatia, 709 Fringilla, 727 Nipponia, 816 Juliana, 909 Ulla, 926 Imhilde, 1128 Astrid, 1189 Terentia, 1222 Tina, 1332 Marconia, 1521 Seinajoki, 2262 Mitidika, 2782 Leonidas, 3556 Lixiaohua, 3811 Karma, 5614 Yakovlev, and 18405 1993FY12.
Three of these families are well studied spectroscopically in the NIR and were also classified as X-complex \citep[i.e. Kalliope, Ursula, Lixiaohua, see][]{avdellidou2025,depra2020a,depra2020b}. 
The SDSS data of Ulla do not match with Gaia DR3, showing a different class. Sylvia, Brasilia and Ursula appear bluer in Gaia DR3 compared to SDSS data, while Aeria appears redder.
The Sylvia family appears slightly redder using the AFP25 catalogue, the family average reflectance spectra of which has T-type as the second-best class, with the SDSS average reflectance spectra agreeing well with the Gaia DR3 average reflectance spectra in AFP25. The Misa family average reflectance spectra in both catalogues appear very similar, but the first-best class changes from X-complex in NES15 to C-complex in AFP25 due to small differences in the average reflectance spectra.
Terentia and Marconia have no SDSS data to perform a comparison. For the rest of the families Gaia DR3 and SDSS agree well.
Checking in both family catalogues, the majority of the families show a uniform $p_\text{V}$ distribution; a few show a potential bimodal distribution (e.g., Aeria, Charis, Mitidika, Konig) and a few appear to have "tails" of just a few objects with higher $p_\text{V}$ values (e.g., Padua, Ursula, Chloris, Croatia, Juliana). These X-complex families consist of dark (e.g., $p_\text{V}<0.1$) members and only five families appear to have medium $p_\text{V}$ values (Kalliope, Brasilia, Tina, Seinajoki, 1993FY12). Medium albedo X-complex asteroids are associated to iron, stony-iron and enstatite chondrite (EH and EL) compositions. Specifically, Kalliope family has been linked to iron/stony-iron meteorites \citep{avdellidou2025}, while the only other medium albedo X-complex family, Athor, that is located in the inner main belt has been uniquely linked to enstatite chondrite meteorites of type EL \citep{avdellidou2022}. Future NIR observations will help in linking these families to meteorites and especially identify the missing source of the EH meteorites.
The 780 Armenia and 15454 1998YB3 families are both classified as X-complex for the first-best class and T-type and C-complex as the second-best class respectively in NES15.
Only two Armenia family members have Gaia DR3 data and their average spectrum is noisy and the one SDSS spectrum available appears to be X-complex. 
In the AFP25 catalogue, one extra member of the Armenia family has Gaia DR3 spectrum, with the average reflectance spectrum having T-type and X-complex as the first- and second-best class respectively. Due to the low number of spectra, all of which are noisy, we do not include this family in our eight reddest families.
With visual inspection and taking into account the SDSS average reflectance spectrum, 1998YB3 appears to be between classes.

\subsubsection{T-/D-type families}
Finally, we report the identification of eight families that appear redder than all the previously mentioned families. 322 Phaeo, 627 Charis, 1303 Luthera, 5438 Lorre and  5567 Durisen are best matched T-types, while 96 Aegle, 1484 Postrema and 53546 2000BY6 with D-types. 
The Postrema family appears bluer in AFP25, with the average reflectance spectrum matching best with the T-types for the first-class. The SDSS and Gaia DR3 average reflectance spectra agree better in AFP25, while in the NES15 catalogue, the SDSS average reflectance spectra is bluer compared to the Gaia DR3 average reflectance spectra.
Lorre is a tiny family with only two members reported in NES15, of which only one has a Gaia DR3 spectrum and none have SDSS data.
Although Lorre consists of 93 members in the AFP25 catalogue, similar to NES15, only the parent asteroid (5438) has a Gaia DR3 spectrum. Hence, the result should be considered with caution.
Similarly, Durisen's Gaia DR3 average reflectance spectrum is built using only five members; however, it generally agrees with the SDSS average reflectance spectrum. In any case, the $p_\text{V}$ of both Lorre and Durisen is very low and does not contradict this classification. Gaia DR3 and SDSS data agree for Phaeo and 2000BY6, while Postrema appears redder in Gaia DR3. There is also a good agreement between the Gaia DR3 and SDSS spectra for Aegle, Luthera and Durisen families, except for the SDSS reflectance at $\approx$0.9 µm which is slightly lower than the values in Gaia DR3 (\autoref{fig:part-average-spectra}). Charis average spectra from the two surveys do not agree, and the family appears to have a highly contaminated $p_\text{V}$ distribution, showing a prominent peak at dark values ($p_\text{V}<0.1$) and a long "tail" at brighter values. 

\subsection{NEA source regions}
\begin{figure}
	\centering
	\includegraphics[width=0.9\linewidth]{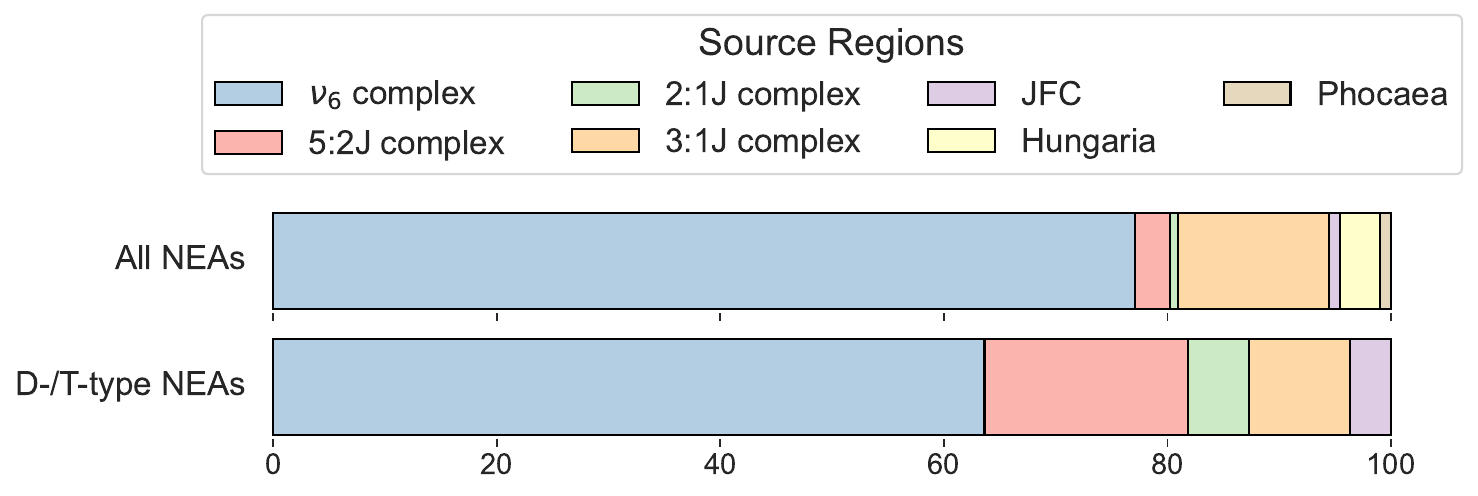}
	\caption{Proportion of NEA source regions. The top bar depicts the breakdown of sources for all NEAs in MP3C, and the bottom bar depicts the breakdown for the T-/D-type NEAs using only spectroscopic observations.}
	\label{fig:nea-source-regions}
\end{figure}

\begin{figure}
	\centering
	\includegraphics[width=0.8\linewidth]{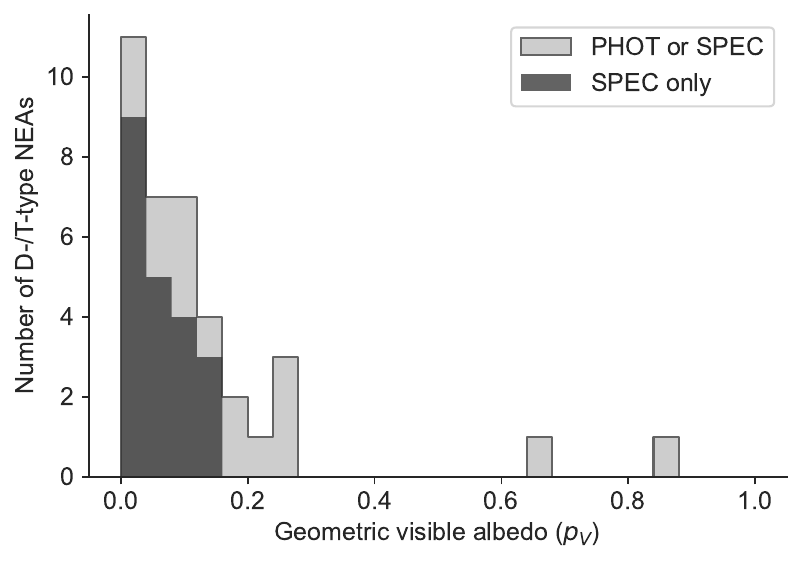}
	\caption{Albedo distribution of T-/D-type NEAs. The dark grey bars represent NEAs that have been classified using either photometric or spectroscopic observations, and the light grey bars represent NEAs that have been classified using only spectroscopic observations.}
	\label{fig:d-nea-albedo}
\end{figure}

To understand if any of the eight reddest families are potential sources for the excess of D-type asteroids in the near-Earth asteroid (NEA) population \citep{perna2018,marsset2022}, we use the following procedure. First, we retrieved the orbits of all NEAs from MP3C and calculated the source region probabilities using the method described in \citet{granvik2018} to get the baseline distribution for the source regions of NEAs. The proportions of NEAs being fed by the different source regions is depicted in \autoref{fig:nea-source-regions}. Next, we selected only the NEAs that have been previously classified as either D- or T-type asteroids in the BDM taxonomy from previous studies using photometric and/or spectroscopic observations. When looking at the albedo distribution of these T-/D-type NEAs (\autoref{fig:d-nea-albedo}), we find that there may be some contamination in the spectroscopic photometric classification that used only photometric data as there are several T-/D-type NEAs with unusual high albedos ($p_\text{V}>0.15$). Thus, from here on, we will only consider NEAs that have been classified using spectroscopic observations. 

\begin{figure*}
	\centering
	\includegraphics[width=\textwidth]{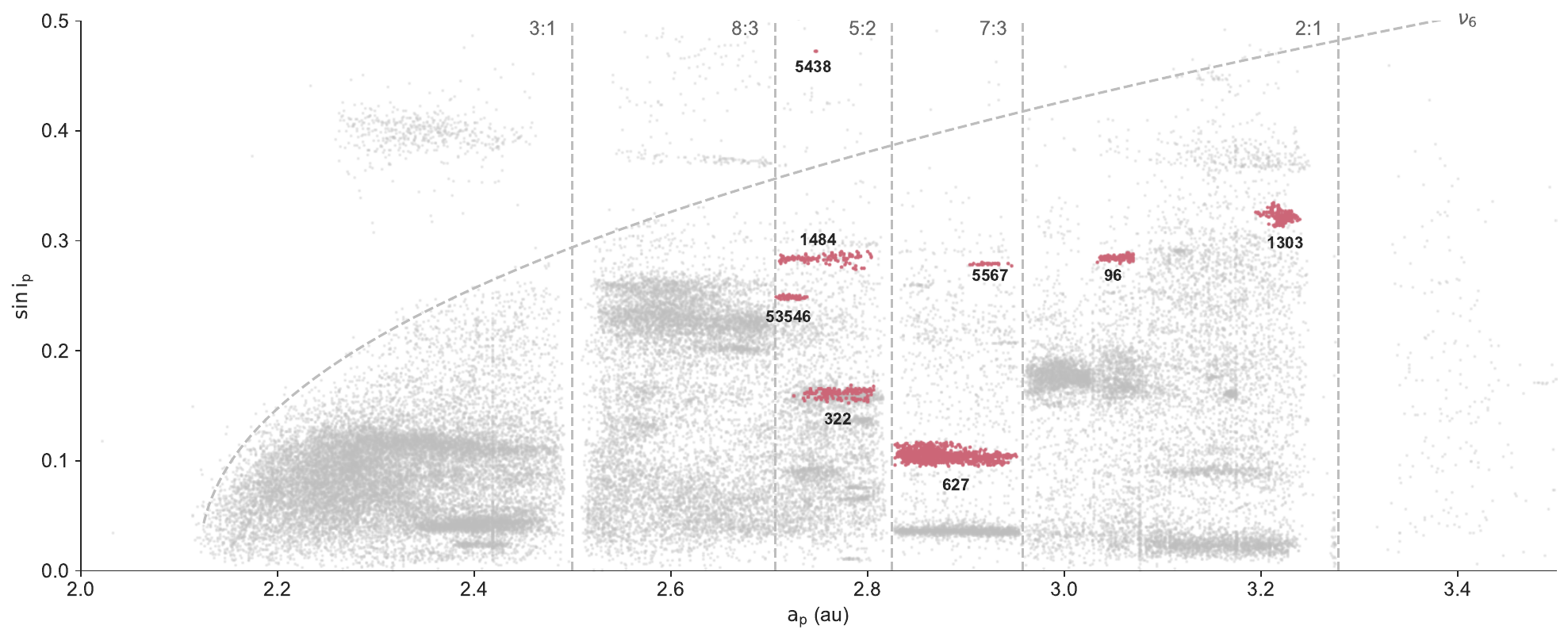}
	\caption{Location of the eight T-/D-type families in the proper semi-major axis vs. proper inclination space. The MMRs with Jupiter and the secular resonance with Saturn are shown as dashed lines. The red points show the very red families (from NES15), with the family number indicated.}
	\label{fig:resonance}
\end{figure*}

The source region breakdown for the T-/D-type NEAs shows an excess of NEAs being fed by the 5:2J-complex and 2:1J-complex resonances \citep[see][for a description of the resonance complexes]{granvik2018} as compared to the general NEA population. Unlike \citet{marsset2022} who identified an abundance from 3:1J instead of 2:1J, we do not observe an increased number of T-/D-type asteroids from the 3:1J resonance, instead we see a reduction as compared to the general NEA population. As seen in \autoref{fig:resonance}, 322 Phaeo, 627 Charis and 1484 Postrema families, identified as the reddest families in this study, are located close to the 5:2J-complex, and 1303 Luthera is close to the 2:1J-complex, possibly contributing to the excess of the T-/D-type NEAs. The 5:2J-complex contains the 8:3J resonance ($\approx$2.7~au) which is close to the 1484 Postrema and 53546 2000BY6 families. This resonance complex also contains the 7:3J which could be fed by 627 Charis and 5567 Durisen.

\subsection{Confidence of the average reflectance spectra}
\label{sec:z-score}
The proportion of family members with Gaia DR3 spectra varies significantly between families. For example, 490 Veritas family has 1294 members in NES15 catalogue, of which 135 have Gaia DR3 spectra, while the similarly sized 668 Dora family, with 1259 members has 248 members with Gaia DR3 spectra. For an infinite population size, the number of samples needed to achieve a certain confidence level and margin of error can be calculated as:
\begin{equation}
	n' = \frac{z^2 \cdot p \cdot (1-p)}{e^2}
	\label{eq:infinite-sample-size}
\end{equation}
where $z$ is the z-score corresponding to the desired confidence level, $p$ is the estimated proportion of an attribute that is present in the population, and $e$ is the desired margin of error. Since the value of $p$ is unknown, we can assume a value of $p=0.5$ which gives us a conservative estimate of the number of samples needed. For a finite population of size $N$, the number of samples needed can be calculated as:
\begin{equation}
	n = \frac{n'}{1 + \frac{z^2 \cdot p \cdot (1-p)}{e^2 \cdot N}}
\end{equation}
where $n'$ is the number of samples needed for an infinite population calculated using \autoref{eq:infinite-sample-size}.


Here we consider the inverse problem, i.e., given a number of samples $n$ from a population size of $N$, what is the confidence level we can achieve for a certain margin of error $e$. Rearranging the above equations, we get:
\begin{equation}
	z = e \cdot {\sqrt{\frac{n \cdot N}{p \cdot (1-p) \cdot (N-n)}}}
\end{equation}

Using this,
we report in \autoref{tab:family-list} the confidence level for the full family average reflectance spectra to be within a margin of error of 10\% of the Gaia DR3 family average reflectance spectra.

\section{Conclusions}
We report that the asteroid families of 96 Aegle, 322 Phaeo, 627 Charis, 1303 Luthera, 1484 Postrema, 5438 Lorre, 5567 Durisen and 53546 2000BY6, appear to be the reddest families of the main belt based on their VIS Gaia DR3 average spectra with five in NES15 (OR 4 in AFP25) of these families are classified as T- or D-types in both the first and second-best class. If this classification is correct, some of these families could be possibly feeding the excess of T-/D-type NEAs escaping the main belt from the 5:2J and 3:2J resonance complexes, with 322 Phaeo, 627 Charis, 1484 Postrema, 5567 Durisen and 53546 2000BY6 feeding the former, and 1303 Luthera feeding the latter. The principal component analysis of the average reflectance spectra of these families show that they are distinct from the C- and X-complex asteroids, however they do not lie in the core D-type asteroid region in the PC1-PC2 space. Future NIR observations of these eight families could help confirm their red nature and their classification, and distinguish between the T- and D-type classifications. 

\section*{Acknowledgements}
UB acknowledges funding from an STFC PhD studentship. MD acknowledges support from the French space agency CNES. CA and MD acknowledge the support of ANR ORIGINS (ANR-18-CE31-0014).  MD and UB acknowledge financial support from the French Programme National de Planetologie (PNP). This work has used data from the European Space Agency (ESA) Gaia mission (\url{https://www.cosmos.esa.int/gaia}), processed by the Gaia Data Processing and Analysis Consortium (DPAC, \url{https://www.cosmos.esa.int/web/gaia/dpac/consortium}). Funding for the DPAC has been provided by national institutions, in particular the institutions participating in the Gaia Multilateral Agreement. This work is based on data provided by the Minor Planet Physical Properties Catalogue (MP3C) of the Observatoire de la C\^ote d'Azur. This research has made use of the Asteroid Families Portal maintained at the Department of Astronomy/University of Belgrade. UB thanks the LSST-DA Data Science Fellowship Program, which is funded by LSST-DA, the Brinson Foundation, the WoodNext Foundation, and the Research Corporation for Science Advancement Foundation; his participation in the program has benefited this work. We thank the reviewer Georgios Tsirvoulis for his comments, allowing us to improve the manuscript.

\section*{Data Availability}
Orbits, albedos and diameters were obtained from the Minor Planet Physical Properties Catalogue (\url{mp3c.oca.eu}). The family memberships were obtained from Asteroid Families Portal (\url{http://asteroids.matf.bg.ac.rs/fam/}). The Gaia DR3 asteroid reflectance spectra were obtained from the archive website \url{https://archives.esac.esa.int/gaia}.

\bibliographystyle{mnras}
\bibliography{references.bib}

@article{carvano2010,
  title = {{{SDSS-based}} Taxonomic Classification and Orbital Distribution of Main Belt Asteroids},
  author = {Carvano, J. M. and Hasselmann, P. H. and Lazzaro, D. and {Moth{\'e}-Diniz}, T.},
  year = {2010},
  month = feb,
  journal = {\aap},
  volume = {510},
  pages = {A43},
  publisher = {EDP},
  issn = {0004-6361},
  doi = {10.1051/0004-6361/200913322},
  urldate = {2025-04-23},
  adsurl = {https://ui.adsabs.harvard.edu/abs/2010A\&A...510A..43C}
}

@article{delbo2017,
  title = {Identification of a Primordial Asteroid Family Constrains the Original Planetesimal Population},
  author = {Delbo', Marco and Walsh, Kevin and Bolin, Bryce and Avdellidou, Chrysa and Morbidelli, Alessandro},
  year = {2017},
  month = sep,
  journal = {Science},
  volume = {357},
  pages = {1026--1029},
  issn = {0036-8075},
  doi = {10.1126/science.aam6036},
  urldate = {2025-03-11},
  adsurl = {https://ui.adsabs.harvard.edu/abs/2017Sci...357.1026D}
}

@inproceedings{delbo2022,
  title = {The {{Minor Planet Physical Properties Catalogue}}: {{Connection}} with the {{Virtual European Solar}} and {{Planetary Access}} of {{EUROPLANET}} and the Big Data Challenge for Planetary Science},
  shorttitle = {The {{Minor Planet Physical Properties Catalogue}}},
  booktitle = {European {{Planetary Science Congress}}},
  author = {Delbo, Marco and Avdellidou, Chrysa and Bruot, Nicolas and Erard, Stephane},
  year = {2022},
  month = sep,
  pages = {EPSC2022-323},
  doi = {10.5194/epsc2022-323},
  urldate = {2025-04-23}
}

@article{demeo2013,
  title = {The Taxonomic Distribution of Asteroids from Multi-Filter All-Sky Photometric Surveys},
  author = {DeMeo, F. E. and Carry, B.},
  year = {2013},
  month = sep,
  journal = {Icarus},
  volume = {226},
  pages = {723--741},
  issn = {0019-1035},
  doi = {10.1016/j.icarus.2013.06.027},
  urldate = {2025-03-19},
  adsurl = {https://ui.adsabs.harvard.edu/abs/2013Icar..226..723D}
}

@article{gaiacollaboration2023,
  title = {Gaia {{Data Release}} 3. {{Reflectance}} Spectra of {{Solar System}} Small Bodies},
  author = {{Gaia Collaboration} and Galluccio, L. and Delbo, M. and De Angeli, F. and Pauwels, T. and Tanga, P. and Mignard, F. and Cellino, A. and Brown, A. G. A. and Muinonen, K. and Penttil{\"a}, A. and Jordan, S. and Vallenari, A. and Prusti, T. and {de Bruijne}, J. H. J. and Arenou, F. and Babusiaux, C. and Biermann, M. and Creevey, O. L. and Ducourant, C. and Evans, D. W. and Eyer, L. and Guerra, R. and Hutton, A. and Jordi, C. and Klioner, S. A. and Lammers, U. L. and Lindegren, L. and Luri, X. and Panem, C. and Pourbaix, D. and Randich, S. and Sartoretti, P. and Soubiran, C. and Walton, N. A. and {Bailer-Jones}, C. A. L. and Bastian, U. and Drimmel, R. and Jansen, F. and Katz, D. and Lattanzi, M. G. and {van Leeuwen}, F. and Bakker, J. and Cacciari, C. and Casta{\~n}eda, J. and Fabricius, C. and Fouesneau, M. and Fr{\'e}mat, Y. and Guerrier, A. and Heiter, U. and Masana, E. and Messineo, R. and Mowlavi, N. and Nicolas, C. and Nienartowicz, K. and Pailler, F. and Panuzzo, P. and Riclet, F. and Roux, W. and Seabroke, G. M. and Sordo, R. and Th{\'e}venin, F. and {Gracia-Abril}, G. and Portell, J. and Teyssier, D. and Altmann, M. and Andrae, R. and Audard, M. and {Bellas-Velidis}, I. and Benson, K. and Berthier, J. and Blomme, R. and Burgess, P. W. and Busonero, D. and Busso, G. and C{\'a}novas, H. and Carry, B. and Cheek, N. and Clementini, G. and Damerdji, Y. and Davidson, M. and {de Teodoro}, P. and Nu{\~n}ez Campos, M. and Delchambre, L. and Dell'Oro, A. and Esquej, P. and {Fern{\'a}ndez-Hern{\'a}ndez}, J. and Fraile, E. and Garabato, D. and {Garc{\'i}a-Lario}, P. and Gosset, E. and Haigron, R. and Halbwachs, J. -L. and Hambly, N. C. and Harrison, D. L. and Hern{\'a}ndez, J. and Hestroffer, D. and Hodgkin, S. T. and Holl, B. and Jan{\ss}en, K. and {Jevardat de Fombelle}, G. and {Krone-Martins}, A. and Lanzafame, A. C. and L{\"o}ffler, W. and Marchal, O. and Marrese, P. M. and Moitinho, A. and Osborne, P. and Pancino, E. and {Recio-Blanco}, A. and Reyl{\'e}, C. and Riello, M. and Rimoldini, L. and Roegiers, T. and Rybizki, J. and Sarro, L. M. and Siopis, C. and Smith, M. and Sozzetti, A. and Utrilla, E. and {van Leeuwen}, M. and Abbas, U. and {\'A}brah{\'a}m, P. and Abreu Aramburu, A. and Aerts, C. and Aguado, J. J. and Ajaj, M. and {Aldea-Montero}, F. and Altavilla, G. and {\'A}lvarez, M. A. and Alves, J. and Anderson, R. I. and Anglada Varela, E. and Antoja, T. and Baines, D. and Baker, S. G. and {Balaguer-N{\'u}{\~n}ez}, L. and Balbinot, E. and Balog, Z. and Barache, C. and Barbato, D. and Barros, M. and Barstow, M. A. and Bartolom{\'e}, S. and Bassilana, J. -L. and Bauchet, N. and Becciani, U. and Bellazzini, M. and Berihuete, A. and Bernet, M. and Bertone, S. and Bianchi, L. and Binnenfeld, A. and {Blanco-Cuaresma}, S. and Boch, T. and Bombrun, A. and Bossini, D. and Bouquillon, S. and Bragaglia, A. and Bramante, L. and Breedt, E. and Bressan, A. and Brouillet, N. and Brugaletta, E. and Bucciarelli, B. and Burlacu, A. and Butkevich, A. G. and Buzzi, R. and Caffau, E. and Cancelliere, R. and {Cantat-Gaudin}, T. and Carballo, R. and Carlucci, T. and Carnerero, M. I. and Carrasco, J. M. and Casamiquela, L. and Castellani, M. and {Castro-Ginard}, A. and Chaoul, L. and Charlot, P. and Chemin, L. and Chiaramida, V. and Chiavassa, A. and Chornay, N. and Comoretto, G. and Contursi, G. and Cooper, W. J. and Cornez, T. and Cowell, S. and Crifo, F. and Cropper, M. and Crosta, M. and Crowley, C. and Dafonte, C. and Dapergolas, A. and David, P. and {de Laverny}, P. and De Luise, F. and De March, R. and De Ridder, J.},
  year = {2023},
  month = jun,
  journal = {\aap},
  volume = {674},
  pages = {A35},
  issn = {0004-6361},
  doi = {10.1051/0004-6361/202243791},
  urldate = {2025-04-24},
  adsurl = {https://ui.adsabs.harvard.edu/abs/2023A\&A...674A..35G}
}

@article{gayon-markt2012,
  title = {On the Origin of the {{Almahata Sitta}} Meteorite and 2008 {{TC3}} Asteroid},
  author = {{Gayon-Markt}, Julie and Delbo, Marco and Morbidelli, Alessandro and Marchi, Simone},
  year = {2012},
  month = jul,
  journal = {\mnras},
  volume = {424},
  pages = {508--518},
  publisher = {OUP},
  issn = {0035-8711},
  doi = {10.1111/j.1365-2966.2012.21220.x},
  urldate = {2025-04-23},
  adsurl = {https://ui.adsabs.harvard.edu/abs/2012MNRAS.424..508G}
}

@inproceedings{ivezic2002,
  title = {Asteroids {{Observed}} by {{The Sloan Digital Sky Survey}}},
  booktitle = {Survey and {{Other Telescope Technologies}} and {{Discoveries}}},
  author = {Ivezic, Zeljko and Juric, M. and Lupton, Robert H. and Tabachnik, S. and Quinn, T.},
  year = {2002},
  month = dec,
  volume = {4836},
  pages = {98--103},
  address = {eprint: arXiv:astro-ph/0208099},
  doi = {10.1117/12.457304},
  urldate = {2025-03-12}
}

@incollection{nesvorny2015,
  title = {Identification and {{Dynamical Properties}} of {{Asteroid Families}}},
  booktitle = {Asteroids {{IV}}},
  author = {Nesvorn{\'y}, D. and Bro{\v z}, M. and Carruba, V.},
  year = {2015},
  month = jan,
  pages = {297--321},
  publisher = {Univesity of Arizona Press},
  doi = {10.2458/azu_uapress_9780816532131-ch016},
  urldate = {2025-03-21}
}

@article{tinaut-ruano2023,
  title = {Asteroids' Reflectance from {{Gaia DR3}}: {{Artificial}} Reddening at near-{{UV}} Wavelengths},
  shorttitle = {Asteroids' Reflectance from {{Gaia DR3}}},
  author = {{Tinaut-Ruano}, F. and Tatsumi, E. and Tanga, P. and {de Le{\'o}n}, J. and Delbo, M. and De Angeli, F. and Morate, D. and Licandro, J. and Galluccio, L.},
  year = {2023},
  month = jan,
  journal = {\aap},
  volume = {669},
  pages = {L14},
  issn = {0004-6361},
  doi = {10.1051/0004-6361/202245134},
  urldate = {2025-04-24},
  adsurl = {https://ui.adsabs.harvard.edu/abs/2023A\&A...669L..14T}
}

@article{delbo2023,
  title = {Gaia View of Primitive Inner-Belt Asteroid Families. {{Searching}} for the Origins of Asteroids {{Bennu}} and {{Ryugu}}},
  author = {Delbo, M. and Avdellidou, C. and Walsh, K. J.},
  year = {2023},
  month = dec,
  journal = {\aap},
  volume = {680},
  pages = {A10},
  issn = {0004-6361},
  doi = {10.1051/0004-6361/202346452},
  urldate = {2025-03-11},
  adsurl = {https://ui.adsabs.harvard.edu/abs/2023A\&A...680A..10D}
}

@article{demeo2009,
  title = {An Extension of the {{Bus}} Asteroid Taxonomy into the Near-Infrared},
  author = {DeMeo, Francesca E. and Binzel, Richard P. and Slivan, Stephen M. and Bus, Schelte J.},
  year = {2009},
  month = jul,
  journal = {Icarus},
  volume = {202},
  pages = {160--180},
  publisher = {Elsevier},
  issn = {0019-1035},
  doi = {10.1016/j.icarus.2009.02.005},
  urldate = {2025-04-24},
  adsurl = {https://ui.adsabs.harvard.edu/abs/2009Icar..202..160D}
}

@article{morate2019,
  title = {The Last Pieces of the Primitive Inner Belt Puzzle: {{Klio}}, {{Chaldaea}}, {{Chimaera}}, and {{Svea}}},
  shorttitle = {The Last Pieces of the Primitive Inner Belt Puzzle},
  author = {Morate, David and {de Le{\'o}n}, Julia and De Pr{\'a}, M{\'a}rio and Licandro, Javier and {Pinilla-Alonso}, Noem{\'i} and Campins, Humberto and Arredondo, Anicia and Carvano, Jorge Marcio and Lazzaro, Daniela and {Cabrera-Lavers}, Antonio},
  year = {2019},
  month = oct,
  journal = {\aap},
  volume = {630},
  pages = {A141},
  publisher = {EDP},
  issn = {0004-6361},
  doi = {10.1051/0004-6361/201935992},
  urldate = {2025-04-24},
  adsurl = {https://ui.adsabs.harvard.edu/abs/2019A\&A...630A.141M}
}

@article{granvik2018,
  title = {Debiased Orbit and Absolute-Magnitude Distributions for near-{{Earth}} Objects},
  author = {Granvik, Mikael and Morbidelli, Alessandro and Jedicke, Robert and Bolin, Bryce and Bottke, William F. and Beshore, Edward and Vokrouhlick{\'y}, David and Nesvorn{\'y}, David and Michel, Patrick},
  year = {2018},
  month = sep,
  journal = {Icarus},
  volume = {312},
  pages = {181--207},
  issn = {0019-1035},
  doi = {10.1016/j.icarus.2018.04.018},
  urldate = {2025-06-06},
  adsurl = {https://ui.adsabs.harvard.edu/abs/2018Icar..312..181G}
}

@article{marsset2022,
  title = {The {{Debiased Compositional Distribution}} of {{MITHNEOS}}: {{Global Match}} between the {{Near-Earth}} and {{Main-belt Asteroid Populations}}, and {{Excess}} of {{D-type Near-Earth Objects}}},
  shorttitle = {The {{Debiased Compositional Distribution}} of {{MITHNEOS}}},
  author = {Marsset, Micha{\"e}l and DeMeo, Francesca E. and Burt, Brian and Polishook, David and Binzel, Richard P. and Granvik, Mikael and Vernazza, Pierre and Carry, Benoit and Bus, Schelte J. and Slivan, Stephen M. and Thomas, Cristina A. and Moskovitz, Nicholas A. and Rivkin, Andrew S.},
  year = {2022},
  month = apr,
  journal = {\aj},
  volume = {163},
  pages = {165},
  publisher = {IOP},
  issn = {0004-6256},
  doi = {10.3847/1538-3881/ac532f},
  urldate = {2025-06-26},
  adsurl = {https://ui.adsabs.harvard.edu/abs/2022AJ....163..165M}
}

@ARTICLE{tsiganis2005,
   author = {{Tsiganis}, K. and {Gomes}, R. and {Morbidelli}, A. and {Levison}, H.~F.
	},
    title = "{Origin of the orbital architecture of the giant planets of the Solar System}",
  journal = {\nat},
     year = 2005,
    month = may,
   volume = 435,
    pages = {459},
      doi = {10.1038/nature03539},
   adsurl = {http://adsabs.harvard.edu/abs/2005Natur.435..459T}
}

@ARTICLE{avdellidou2024,
       author = {{Avdellidou}, Chrysa and {Delbo{\textquoteright}}, Marco and {Nesvorn{\'y}}, David and {Walsh}, Kevin J. and {Morbidelli}, Alessandro},
        title = "{Dating the Solar System{\textquoteright}s giant planet orbital instability using enstatite meteorites}",
      journal = {Science},
         year = 2024,
        month = apr,
       volume = {384},
       number = {6693},
        pages = {348-352},
          doi = {10.1126/science.adg8092},
       adsurl = {https://ui.adsabs.harvard.edu/abs/2024Sci...384..348A}
}

@ARTICLE{levison2009,
       author = {{Levison}, Harold F. and {Bottke}, William F. and {Gounelle}, Matthieu and {Morbidelli}, Alessandro and {Nesvorn{\'y}}, David and {Tsiganis}, Kleomenis},
        title = "{Contamination of the asteroid belt by primordial trans-Neptunian objects}",
      journal = {\nat},
         year = 2009,
        month = jul,
       volume = {460},
       number = {7253},
        pages = {364-366},
          doi = {10.1038/nature08094},
       adsurl = {https://ui.adsabs.harvard.edu/abs/2009Natur.460..364L}
}

@article{trieloff2022,
title = {Evolution of the parent body of enstatite (EL) chondrites},
journal = {\icarus},
volume = {373},
pages = {114762},
year = {2022},
issn = {0019-1035},
doi = {https://doi.org/10.1016/j.icarus.2021.114762},
url = {https://www.sciencedirect.com/science/article/pii/S0019103521004140},
author = {Mario Trieloff and Jens Hopp and Hans-Peter Gail}
}

@INCOLLECTION{reddy2015,
       author = {{Reddy}, V. and {Dunn}, T.~L. and {Thomas}, C.~A. and {Moskovitz}, N.~A. and {Burbine}, T.~H.},
        title = "{Mineralogy and Surface Composition of Asteroids}",
    booktitle = {Asteroids IV},
    publisher = {University of Arizona Press},
         year = 2015,
       editor = {{Michel}, Patrick and {DeMeo}, Francesca E. and {Bottke}, William F.},
        pages = {43-63},
          doi = {10.2458/azu_uapress_9780816532131-ch003},
       adsurl = {https://ui.adsabs.harvard.edu/abs/2015aste.book...43R}
}

@ARTICLE{deleon2016,
       author = {{de Le{\'o}n}, J. and {Pinilla-Alonso}, N. and {Delbo}, M. and {Campins}, H. and {Cabrera-Lavers}, A. and {Tanga}, P. and {Cellino}, A. and {Bendjoya}, P. and {Gayon-Markt}, J. and {Licandro}, J. and {Lorenzi}, V. and {Morate}, D. and {Walsh}, K.~J. and {DeMeo}, F. and {Landsman}, Z. and {Al{\'\i}-Lagoa}, V.},
        title = "{Visible spectroscopy of the Polana-Eulalia family complex: Spectral homogeneity}",
      journal = {\icarus},
         year = 2016,
        month = mar,
       volume = {266},
        pages = {57-75},
          doi = {10.1016/j.icarus.2015.11.014},
       adsurl = {https://ui.adsabs.harvard.edu/abs/2016Icar..266...57D}
}

@ARTICLE{pinilla_alonso2016,
       author = {{Pinilla-Alonso}, Noem{\'\i} and {de Le{\'o}n}, J. and {Walsh}, K.~J. and {Campins}, H. and {Lorenzi}, V. and {Delbo}, M. and {DeMeo}, F. and {Licandro}, J. and {Landsman}, Z. and {Lucas}, M.~P. and {Al{\'\i}-Lagoa}, V. and {Burt}, B.},
        title = "{Portrait of the Polana-Eulalia family complex: Surface homogeneity revealed from near-infrared spectroscopy}",
      journal = {\icarus},
         year = 2016,
        month = aug,
       volume = {274},
        pages = {231-248},
          doi = {10.1016/j.icarus.2016.03.022},
       adsurl = {https://ui.adsabs.harvard.edu/abs/2016Icar..274..231P}
}

@ARTICLE{morate2016,
       author = {{Morate}, David and {de Le{\'o}n}, Julia and {De Pr{\'a}}, M{\'a}rio and {Licandro}, Javier and {Cabrera-Lavers}, Antonio and {Campins}, Humberto and {Pinilla-Alonso}, Noem{\'\i} and {Al{\'\i}-Lagoa}, V{\'\i}ctor},
        title = "{Compositional study of asteroids in the Erigone collisional family using visible spectroscopy at the 10.4 m GTC}",
      journal = {\aap},
         year = 2016,
        month = feb,
       volume = {586},
          eid = {A129},
        pages = {A129},
          doi = {10.1051/0004-6361/201527453},
archivePrefix = {arXiv},
       eprint = {1701.03761},
 primaryClass = {astro-ph.EP},
       adsurl = {https://ui.adsabs.harvard.edu/abs/2016A&A...586A.129M}
}

@ARTICLE{morate2018,
       author = {{Morate}, David and {de Le{\'o}n}, Julia and {De Pr{\'a}}, M{\'a}rio and {Licandro}, Javier and {Cabrera-Lavers}, Antonio and {Campins}, Humberto and {Pinilla-Alonso}, Noem{\'\i}},
        title = "{Visible spectroscopy of the Sulamitis and Clarissa primitive families: a possible link to Erigone and Polana}",
      journal = {\aap},
         year = 2018,
        month = feb,
       volume = {610},
          eid = {A25},
        pages = {A25},
          doi = {10.1051/0004-6361/201731407},
       adsurl = {https://ui.adsabs.harvard.edu/abs/2018A&A...610A..25M}
}

@ARTICLE{avdellidou2025,
       author = {{Avdellidou}, Chrysa and {Bhat}, Ullas and {Bujdoso}, Kieran and {Delbo}, Marco and {Marsset}, Michael and {Vernazza}, Pierre},
        title = "{Kalliope sings rock and metal}",
      journal = {\mnras},
         year = 2025,
        month = jun,
       volume = {539},
       number = {4},
        pages = {3534-3550},
          doi = {10.1093/mnras/staf640},
       adsurl = {https://ui.adsabs.harvard.edu/abs/2025MNRAS.539.3534A}
}

@ARTICLE{migliorini1996,
       author = {{Migliorini}, F. and {Manara}, A. and {di Martino}, M. and {Farinella}, P.},
        title = "{The Hoffmeister asteroid family: inferences from physical data.}",
      journal = {\aap},
         year = 1996,
        month = jun,
       volume = {310},
        pages = {681-685},
       adsurl = {https://ui.adsabs.harvard.edu/abs/1996A&A...310..681M}
}

@ARTICLE{fornasier2016,
       author = {{Fornasier}, S. and {Lantz}, C. and {Perna}, D. and {Campins}, H. and {Barucci}, M.~A. and {Nesvorny}, D.},
        title = "{Spectral variability on primitive asteroids of the Themis and Beagle families: Space weathering effects or parent body heterogeneity?}",
      journal = {\icarus},
         year = 2016,
        month = may,
       volume = {269},
        pages = {1-14},
          doi = {10.1016/j.icarus.2016.01.002},
archivePrefix = {arXiv},
       eprint = {1601.05277},
 primaryClass = {astro-ph.EP},
       adsurl = {https://ui.adsabs.harvard.edu/abs/2016Icar..269....1F}
}

@ARTICLE{marsset2016,
   author = {{Marsset}, M. and {Vernazza}, P. and {Birlan}, M. and {DeMeo}, F. and 
	{Binzel}, R.~P. and {Dumas}, C. and {Milli}, J. and {Popescu}, M.
	},
    title = "{Compositional characterisation of the Themis family}",
  journal = {\aap},
archivePrefix = "arXiv",
   eprint = {1601.02405},
 primaryClass = "astro-ph.EP",
     year = 2016,
    month = feb,
   volume = 586,
      eid = {A15},
    pages = {A15},
      doi = {10.1051/0004-6361/201526962},
   adsurl = {http://adsabs.harvard.edu/abs/2016A%26A...586A..15M}
}

@ARTICLE{yang2020,
       author = {{Yang}, B. and {Hanu{\v{s}}}, J. and {Bro{\v{z}}}, M. and {Chrenko}, O. and {Willman}, M. and {{\v{S}}eve{\v{c}}ek}, P. and {Masiero}, J. and {Kaluna}, H.},
        title = "{Physical and dynamical characterization of the Euphrosyne asteroid family}",
      journal = {\aap},
         year = 2020,
        month = nov,
       volume = {643},
          eid = {A38},
        pages = {A38},
          doi = {10.1051/0004-6361/202038567},
archivePrefix = {arXiv},
       eprint = {2009.04489},
 primaryClass = {astro-ph.EP},
       adsurl = {https://ui.adsabs.harvard.edu/abs/2020A&A...643A..38Y}
}

@ARTICLE{depra2020a,
       author = {{De Pr{\'a}}, M.~N. and {Licandro}, J. and {Pinilla-Alonso}, N. and {Lorenzi}, V. and {Rond{\'o}n}, E. and {Carvano}, J. and {Morate}, D. and {De Le{\'o}n}, J.},
        title = "{The spectroscopic properties of the Lixiaohua family, cradle of Main Belt Comets}",
      journal = {\icarus},
         year = 2020,
        month = mar,
       volume = {338},
          eid = {113473},
        pages = {113473},
          doi = {10.1016/j.icarus.2019.113473},
       adsurl = {https://ui.adsabs.harvard.edu/abs/2020Icar..33813473D}
}

@ARTICLE{depra2020b,
       author = {{De Pr{\'a}}, M.~N. and {Pinilla-Alonso}, N. and {Carvano}, J. and {Licandro}, J. and {Morate}, D. and {Lorenzi}, V. and {de Le{\'o}n}, J. and {Campins}, H. and {Moth{\'e}-Diniz}, T.},
        title = "{A comparative analysis of the outer-belt primitive families}",
      journal = {\aap},
         year = 2020,
        month = nov,
       volume = {643},
          eid = {A102},
        pages = {A102},
          doi = {10.1051/0004-6361/202038536},
archivePrefix = {arXiv},
       eprint = {2105.11994},
 primaryClass = {astro-ph.EP},
       adsurl = {https://ui.adsabs.harvard.edu/abs/2020A&A...643A.102D}
}

@ARTICLE{avdellidou2022,
       author = {{Avdellidou}, C. and {Delbo}, M. and {Morbidelli}, A. and {Walsh}, K.~J. and {Munaibari}, E. and {Bourdelle de Micas}, J. and {Devog{\`e}le}, M. and {Fornasier}, S. and {Gounelle}, M. and {van Belle}, G.},
        title = "{Athor asteroid family as the source of the EL enstatite meteorites}",
      journal = {\aap},
         year = 2022,
        month = sep,
       volume = {665},
          eid = {L9},
        pages = {L9},
          doi = {10.1051/0004-6361/202244590},
       adsurl = {https://ui.adsabs.harvard.edu/abs/2022A&A...665L...9A}
}

@ARTICLE{arredondo2020,
       author = {{Arredondo}, Anicia and {Lorenzi}, Vania and {Pinilla-Alonso}, Noemi and {Campins}, Humberto and {Malfavon}, Andrew and {de Le{\'o}n}, Julia and {Morate}, David},
        title = "{Near-infrared spectroscopy of the Klio primitive inner-belt asteroid family}",
      journal = {\icarus},
         year = 2020,
        month = jan,
       volume = {335},
          eid = {113427},
        pages = {113427},
          doi = {10.1016/j.icarus.2019.113427},
       adsurl = {https://ui.adsabs.harvard.edu/abs/2020Icar..33513427A}
}

@ARTICLE{arredondo2021b,
       author = {{Arredondo}, Anicia and {Campins}, Humberto and {Pinilla-Alonso}, Noemi and {de Le{\'o}n}, Julia and {Lorenzi}, Vania and {Morate}, David},
        title = "{Near-infrared spectroscopy of the Sulamitis asteroid family: Surprising similarities in the inner belt primitive asteroid population}",
      journal = {\icarus},
         year = 2021,
        month = apr,
       volume = {358},
          eid = {114210},
        pages = {114210},
          doi = {10.1016/j.icarus.2020.114210},
       adsurl = {https://ui.adsabs.harvard.edu/abs/2021Icar..35814210A}
}

@ARTICLE{delbo2019,
       author = {{Delbo}, Marco and {Avdellidou}, Chrysa and {Morbidelli}, Alessandro},
        title = "{Ancient and primordial collisional families as the main sources of X-type asteroids of the inner main belt}",
      journal = {\aap},
         year = 2019,
        month = apr,
       volume = {624},
          eid = {A69},
        pages = {A69},
          doi = {10.1051/0004-6361/201834745},
archivePrefix = {arXiv},
       eprint = {1902.01633},
 primaryClass = {astro-ph.EP},
       adsurl = {https://ui.adsabs.harvard.edu/abs/2019A&A...624A..69D}
}

@ARTICLE{broz2022,
       author = {{Bro{\v{z}}}, M. and {Ferrais}, M. and {Vernazza}, P. and {{\v{S}}eve{\v{c}}ek}, P. and {Jutzi}, M.},
        title = "{Discovery of an asteroid family linked to (22) Kalliope and its moon Linus}",
      journal = {\aap},
         year = 2022,
        month = aug,
       volume = {664},
          eid = {A69},
        pages = {A69},
          doi = {10.1051/0004-6361/202243628},
archivePrefix = {arXiv},
       eprint = {2205.15736},
 primaryClass = {astro-ph.EP},
       adsurl = {https://ui.adsabs.harvard.edu/abs/2022A&A...664A..69B}
}

@ARTICLE{demeo2022,
       author = {{DeMeo}, Francesca E. and {Burt}, Brian J. and {Marsset}, Micha{\"e}l and {Polishook}, David and {Burbine}, Thomas H. and {Carry}, Beno{\^\i}t and {Binzel}, Richard P. and {Vernazza}, Pierre and {Reddy}, Vishnu and {Tang}, Michelle and {Thomas}, Cristina A. and {Rivkin}, Andrew S. and {Moskovitz}, Nicholas A. and {Slivan}, Stephen M. and {Bus}, Schelte J.},
        title = "{Connecting asteroids and meteorites with visible and near-infrared spectroscopy}",
      journal = {\icarus},
         year = 2022,
        month = jul,
       volume = {380},
          eid = {114971},
        pages = {114971},
          doi = {10.1016/j.icarus.2022.114971},
archivePrefix = {arXiv},
       eprint = {2202.13797},
 primaryClass = {astro-ph.EP},
       adsurl = {https://ui.adsabs.harvard.edu/abs/2022Icar..38014971D}
}

@ARTICLE{vernazza2009,
   author = {{Vernazza}, P. and {Binzel}, R.~P. and {Rossi}, A. and {Fulchignoni}, M. and 
	{Birlan}, M.},
    title = "{Solar wind as the origin of rapid reddening of asteroid surfaces}",
  journal = {\nat},
     year = 2009,
    month = apr,
   volume = 458,
    pages = {993},
      doi = {10.1038/nature07956},
   adsurl = {http://adsabs.harvard.edu/abs/2009Natur.458..993V}
}

@ARTICLE{gaffey1992,
       author = {{Gaffey}, Michael J. and {Reed}, Kevin L. and {Kelley}, Michael S.},
        title = "{Relationship of E-type Apollo asteroid 3103 (1982 BB) to the enstatite achondrite meteorites and the Hungaria asteroids}",
      journal = {\icarus},
         year = 1992,
        month = nov,
       volume = {100},
       number = {1},
        pages = {95-109},
          doi = {10.1016/0019-1035(92)90021-X},
       adsurl = {https://ui.adsabs.harvard.edu/abs/1992Icar..100...95G}
}

@ARTICLE{cloutis1990,
       author = {{Cloutis}, E.~A. and {Gaffey}, M.~J. and {Smith}, D.~G.~W. and {Lambert}, R. St. J.},
        title = "{Metal Silicate Mixtures: Spectral Properties and Applications to Asteroid Taxonomy}",
      journal = {\jgr},
         year = 1990,
        month = jun,
       volume = {95},
        pages = {8323-8338},
          doi = {10.1029/JB095iB06p08323},
       adsurl = {https://ui.adsabs.harvard.edu/abs/1990JGR....95.8323C}
}

@article{demeo2015,
   author = {{DeMeo}, F.~E. and {Alexander}, C.~M.~O. and {Walsh}, K.~J. and 
	{Chapman}, C.~R. and {Binzel}, R.~P.},
    title = "{The Compositional Structure of the Asteroid Belt}",
   journal = {in Asteroids IV (P. Michel, et al. eds)},
    editor = {{Michel}, P. and {DeMeo}, F.~E. and {Bottke}, W.~F.},
    pages = {13-41},
     year = 2015,
   adsurl = {http://adsabs.harvard.edu/abs/2015arXiv150604805D}
}

@ARTICLE{jenniskens2025,
       author = {{Jenniskens}, Peter and {Devillepoix}, Hadrien A.~R.},
        title = "{Review of asteroid, meteor, and meteorite-type links}",
      journal = {\maps},
         year = 2025,
        month = apr,
       volume = {60},
       number = {4},
        pages = {928-973},
          doi = {10.1111/maps.14321},
       adsurl = {https://ui.adsabs.harvard.edu/abs/2025M&PS...60..928J}
}

@ARTICLE{russel2012,
   author = {{Russell}, C.~T. and {Raymond}, C.~A. and {Coradini}, A. and 
	{McSween}, H.~Y. and {Zuber}, M.~T. and {Nathues}, A. and {De Sanctis}, M.~C. and 
	{Jaumann}, R. and {Konopliv}, A.~S. and {Preusker}, F. and {Asmar}, S.~W. and 
	{Park}, R.~S. and {Gaskell}, R. and {Keller}, H.~U. and {Mottola}, S. and 
	{Roatsch}, T. and {Scully}, J.~E.~C. and {Smith}, D.~E. and 
	{Tricarico}, P. and {Toplis}, M.~J. and {Christensen}, U.~R. and 
	{Feldman}, W.~C. and {Lawrence}, D.~J. and {McCoy}, T.~J. and 
	{Prettyman}, T.~H. and {Reedy}, R.~C. and {Sykes}, M.~E. and 
	{Titus}, T.~N.},
    title = "{Dawn at Vesta: Testing the Protoplanetary Paradigm}",
  journal = {Science},
     year = 2012,
    month = may,
   volume = 336,
    pages = {684},
      doi = {10.1126/science.1219381},
   adsurl = {http://adsabs.harvard.edu/abs/2012Sci...336..684R}
}

@ARTICLE{lucas2019,
       author = {{Lucas}, Michael P. and {Emery}, Joshua P. and {MacLennan}, Eric M. and {Pinilla-Alonso}, Noemi and {Cartwright}, Richard J. and {Lindsay}, Sean S. and {Reddy}, Vishnu and {Sanchez}, Juan A. and {Thomas}, Cristina A. and {Lorenzi}, Vania},
        title = "{Hungaria asteroid region telescopic spectral survey (HARTSS) II: Spectral homogeneity among Hungaria family asteroids}",
      journal = {\icarus},
         year = 2019,
        month = apr,
       volume = {322},
        pages = {227-250},
          doi = {10.1016/j.icarus.2018.12.010},
       adsurl = {https://ui.adsabs.harvard.edu/abs/2019Icar..322..227L}
}

@ARTICLE{marsset2024,
       author = {{Marsset}, M. and {Vernazza}, P. and {Bro{\v{z}}}, M. and {Thomas}, C.~A. and {DeMeo}, F.~E. and {Burt}, B. and {Binzel}, R.~P. and {Reddy}, V. and {McGraw}, A. and {Avdellidou}, C. and {Carry}, B. and {Slivan}, S. and {Polishook}, D.},
        title = "{The Massalia asteroid family as the origin of ordinary L chondrites}",
      journal = {\nat},
         year = 2024,
        month = oct,
       volume = {634},
       number = {8034},
        pages = {561-565},
          doi = {10.1038/s41586-024-08007-6},
archivePrefix = {arXiv},
       eprint = {2403.08548},
 primaryClass = {astro-ph.EP},
       adsurl = {https://ui.adsabs.harvard.edu/abs/2024Natur.634..561M}
}

@ARTICLE{vernazza2008,
       author = {{Vernazza}, P. and {Binzel}, R.~P. and {Thomas}, C.~A. and {DeMeo}, F.~E. and {Bus}, S.~J. and {Rivkin}, A.~S. and {Tokunaga}, A.~T.},
        title = "{Compositional differences between meteorites and near-Earth asteroids}",
      journal = {\nat},
         year = 2008,
        month = aug,
       volume = {454},
       number = {7206},
        pages = {858-860},
          doi = {10.1038/nature07154},
       adsurl = {https://ui.adsabs.harvard.edu/abs/2008Natur.454..858V}
}

@ARTICLE{fornasier2007,
       author = {{Fornasier}, S. and {Dotto}, E. and {Hainaut}, O. and {Marzari}, F. and {Boehnhardt}, H. and {De Luise}, F. and {Barucci}, M.~A.},
        title = "{Visible spectroscopic and photometric survey of Jupiter Trojans: Final results on dynamical families}",
      journal = {\icarus},
         year = 2007,
        month = oct,
       volume = {190},
       number = {2},
        pages = {622-642},
          doi = {10.1016/j.icarus.2007.03.033},
archivePrefix = {arXiv},
       eprint = {0704.0350},
 primaryClass = {astro-ph},
       adsurl = {https://ui.adsabs.harvard.edu/abs/2007Icar..190..622F}
}

@ARTICLE{avdellidou2021b,
       author = {{Avdellidou}, C. and {Pajola}, M. and {Lucchetti}, A. and {Agostini}, L. and {Delbo}, M. and {Mazzotta Epifani}, E. and {Bourdelle de Micas}, J. and {Devog{\`e}le}, M. and {Fornasier}, S. and {van Belle}, G. and {Bruot}, N. and {Dotto}, E. and {Ieva}, S. and {Cremonese}, G. and {Palumbo}, P.},
        title = "{Characterisation of the main belt asteroid (223) Rosa. A proposed flyby target of ESA's JUICE mission}",
      journal = {\aap},
         year = 2021,
        month = dec,
       volume = {656},
          eid = {L18},
        pages = {L18},
          doi = {10.1051/0004-6361/202142600},
       adsurl = {https://ui.adsabs.harvard.edu/abs/2021A&A...656L..18A}
}

@ARTICLE{hasegawa2022,
       author = {{Hasegawa}, Sunao and {DeMeo}, Francesca E. and {Marsset}, Micha{\"e}l and {Hanu{\v{s}}}, Josef and {Avdellidou}, Chrysa and {Delbo}, Marco and {Bus}, Schelte J. and {Hanayama}, Hidekazu and {Horiuchi}, Takashi and {Takir}, Driss and {Jehin}, Emmanu{\"e}l and {Ferrais}, Marin and {Geem}, Jooyeon and {Im}, Myungshin and {Seo}, Jinguk and {Bach}, Yoonsoo P. and {Jin}, Sunho and {Ishiguro}, Masateru and {Kuroda}, Daisuke and {Binzel}, Richard P. and {Nakamura}, Akiko M. and {Yang}, Bin and {Vernazza}, Pierre},
        title = "{Spectral Evolution of Dark Asteroid Surfaces Induced by Space Weathering over a Decade}",
      journal = {\apj},
         year = 2022,
        month = nov,
       volume = {939},
       number = {1},
          eid = {L9},
        pages = {L9},
          doi = {10.3847/2041-8213/ac92e4},
archivePrefix = {arXiv},
       eprint = {2209.09415},
 primaryClass = {astro-ph.EP},
       adsurl = {https://ui.adsabs.harvard.edu/abs/2022ApJ...939L...9H}
}

@ARTICLE{perna2018,
       author = {{Perna}, D. and {Barucci}, M.~A. and {Fulchignoni}, M. and {Popescu}, M. and {Belskaya}, I. and {Fornasier}, S. and {Doressoundiram}, A. and {Lantz}, C. and {Merlin}, F.},
        title = "{A spectroscopic survey of the small near-Earth asteroid population: Peculiar taxonomic distribution and phase reddening}",
      journal = {\planss},
         year = 2018,
        month = aug,
       volume = {157},
        pages = {82-95},
          doi = {10.1016/j.pss.2018.03.008},
archivePrefix = {arXiv},
       eprint = {1803.08953},
 primaryClass = {astro-ph.EP},
       adsurl = {https://ui.adsabs.harvard.edu/abs/2018P&SS..157...82P}
}

@ARTICLE{demeo2014,
       author = {{DeMeo}, Francesca E. and {Binzel}, Richard P. and {Carry}, Beno{\^\i}t and {Polishook}, David and {Moskovitz}, Nicholas A.},
        title = "{Unexpected D-type interlopers in the inner main belt}",
      journal = {\icarus},
         year = 2014,
        month = feb,
       volume = {229},
        pages = {392-399},
          doi = {10.1016/j.icarus.2013.11.026},
archivePrefix = {arXiv},
       eprint = {1312.2962},
 primaryClass = {astro-ph.EP},
       adsurl = {https://ui.adsabs.harvard.edu/abs/2014Icar..229..392D}
}

@article{virtanen2020,
  title = {{{SciPy}} 1.0: Fundamental Algorithms for Scientific Computing in {{Python}}},
  shorttitle = {{{SciPy}} 1.0},
  author = {Virtanen, Pauli and Gommers, Ralf and Oliphant, Travis E. and Haberland, Matt and Reddy, Tyler and Cournapeau, David and Burovski, Evgeni and Peterson, Pearu and Weckesser, Warren and Bright, Jonathan and {van der Walt}, St{\'e}fan J. and Brett, Matthew and Wilson, Joshua and Millman, K. Jarrod and Mayorov, Nikolay and Nelson, Andrew R. J. and Jones, Eric and Kern, Robert and Larson, Eric and Carey, C. J. and Polat, {\.I}lhan and Feng, Yu and Moore, Eric W. and VanderPlas, Jake and Laxalde, Denis and Perktold, Josef and Cimrman, Robert and Henriksen, Ian and Quintero, E. A. and Harris, Charles R. and Archibald, Anne M. and Ribeiro, Ant{\^o}nio H. and Pedregosa, Fabian and {van Mulbregt}, Paul and {SciPy 1. 0 Contributors}},
  year = {2020},
  month = feb,
  journal = {Nature Methods},
  volume = {17},
  pages = {261--272},
  doi = {10.1038/s41592-019-0686-2},
  urldate = {2025-08-05},
  adsurl = {https://ui.adsabs.harvard.edu/abs/2020NatMe..17..261V}
}

@article{ebel2011,
  title = {Equilibrium Condensation from Chondritic Porous {{IDP}} Enriched Vapor: {{Implications}} for {{Mercury}} and Enstatite Chondrite Origins},
  shorttitle = {Equilibrium Condensation from Chondritic Porous {{IDP}} Enriched Vapor},
  author = {Ebel, D. S. and Alexander, C. M. O'D.},
  year = {2011},
  month = dec,
  journal = {Planetary and Space Science},
  volume = {59},
  pages = {1888--1894},
  issn = {0032-0633},
  doi = {10.1016/j.pss.2011.07.017},
  urldate = {2025-08-26},
  adsurl = {https://ui.adsabs.harvard.edu/abs/2011P\&SS...59.1888E}
}

@incollection{nittler2018,
  title = {The {{Chemical Composition}} of {{Mercury}}},
  booktitle = {Mercury. {{The View}} after {{MESSENGER}}},
  author = {Nittler, Larry R. and Chabot, Nancy L. and Grove, Timothy L. and Peplowski, Patrick N.},
  year = {2018},
  month = jan,
  pages = {30--51},
  publisher = {Cambridge University Press},
  doi = {10.1017/9781316650684.003},
  urldate = {2025-08-26},
  adsurl = {https://ui.adsabs.harvard.edu/abs/2018mvam.book...30N}
}

@article{dauphas2017,
  title = {The Isotopic Nature of the {{Earth}}'s Accreting Material through Time},
  author = {Dauphas, Nicolas},
  year = {2017},
  month = jan,
  journal = {Nature},
  volume = {541},
  pages = {521--524},
  issn = {0028-0836},
  doi = {10.1038/nature20830},
  urldate = {2025-08-26},
  adsurl = {https://ui.adsabs.harvard.edu/abs/2017Natur.541..521D}
}

@article{javoy2010,
  title = {The Chemical Composition of the {{Earth}}: {{Enstatite}} Chondrite Models},
  shorttitle = {The Chemical Composition of the {{Earth}}},
  author = {Javoy, M. and Kaminski, E. and Guyot, F. and Andrault, D. and Sanloup, C. and Moreira, M. and Labrosse, S. and Jambon, A. and Agrinier, P. and Davaille, A. and Jaupart, C.},
  year = {2010},
  month = may,
  journal = {E\&PSL},
  volume = {293},
  pages = {259--268},
  issn = {0012-821X},
  doi = {10.1016/j.epsl.2010.02.033},
  urldate = {2025-08-26},
  adsurl = {https://ui.adsabs.harvard.edu/abs/2010E\&PSL.293..259J}
}

@article{warren2011,
  title = {Stable-Isotopic Anomalies and the Accretionary Assemblage of the {{Earth}} and {{Mars}}: {{A}} Subordinate Role for Carbonaceous Chondrites},
  shorttitle = {Stable-Isotopic Anomalies and the Accretionary Assemblage of the {{Earth}} and {{Mars}}},
  author = {Warren, Paul H.},
  year = {2011},
  month = nov,
  journal = {E\&PSL},
  volume = {311},
  pages = {93--100},
  issn = {0012-821X},
  doi = {10.1016/j.epsl.2011.08.047},
  urldate = {2025-06-25},
  adsurl = {https://ui.adsabs.harvard.edu/abs/2011E\&PSL.311...93W}
}

@article{brown2000,
  title = {The {{Fall}}, {{Recovery}}, {{Orbit}}, and {{Composition}} of the {{Tagish Lake Meteorite}}: {{A New Type}} of {{Carbonaceous Chondrite}}},
  shorttitle = {The {{Fall}}, {{Recovery}}, {{Orbit}}, and {{Composition}} of the {{Tagish Lake Meteorite}}},
  author = {Brown, Peter G. and Hildebrand, Alan R. and Zolensky, Michael E. and Grady, Monica and Clayton, Robert N. and Mayeda, Toshiko K. and Tagliaferri, Edward and Spalding, Richard and MacRae, Neil D. and Hoffman, Eric L. and Mittlefehldt, David W. and Wacker, John F. and Bird, J. Andrew and Campbell, Margaret D. and Carpenter, Robert and Gingerich, Heather and Glatiotis, Michael and Greiner, Erika and Mazur, Michael J. and McCausland, Phil JA. and Plotkin, Howard and Rubak Mazur, Tina},
  year = {2000},
  month = oct,
  journal = {Science},
  volume = {290},
  pages = {320--325},
  issn = {0036-8075},
  doi = {10.1126/science.290.5490.320},
  urldate = {2025-08-28},
  adsurl = {https://ui.adsabs.harvard.edu/abs/2000Sci...290..320B}
}

@article{hiroi2001,
  title = {The {{Tagish Lake Meteorite}}: {{A Possible Sample}} from a {{D-Type Asteroid}}},
  shorttitle = {The {{Tagish Lake Meteorite}}},
  author = {Hiroi, Takahiro and Zolensky, Michael E. and Pieters, Carle M.},
  year = {2001},
  month = sep,
  journal = {Science},
  volume = {293},
  pages = {2234--2236},
  issn = {0036-8075},
  doi = {10.1126/science.1063734},
  urldate = {2025-08-28},
  adsurl = {https://ui.adsabs.harvard.edu/abs/2001Sci...293.2234H}
}

@inproceedings{hiroi2005,
  title = {Meteorite {{WIS91600}}: {{A New Sample Related}} to a {{D-}} or {{T-type Asteroid}}},
  shorttitle = {Meteorite {{WIS91600}}},
  booktitle = {36th {{Annual Lunar}} and {{Planetary Science Conference}}},
  author = {Hiroi, T. and Tonui, E. and Pieters, C. M. and Zolensky, M. E. and Ueda, Y. and Miyamoto, M. and Sasaki, S.},
  year = {2005},
  month = mar,
  pages = {1564},
  urldate = {2025-08-28}
}

@article{marrocchi2021,
  title = {The {{Tarda Meteorite}}: {{A Window}} into the {{Formation}} of {{D-type Asteroids}}},
  shorttitle = {The {{Tarda Meteorite}}},
  author = {Marrocchi, Yves and Avice, Guillaume and Barrat, Jean-Alix},
  year = {2021},
  month = may,
  journal = {\apj},
  volume = {913},
  pages = {L9},
  issn = {0004-637X},
  doi = {10.3847/2041-8213/abfaa3},
  urldate = {2025-08-28},
  adsurl = {https://ui.adsabs.harvard.edu/abs/2021ApJ...913L...9M}
}

@article{nakamura2013,
  title = {{{MET}} 00432: {{Another Tagish Lake-Type Carbonaceous Chondrite}} from {{Antarctica}}},
  shorttitle = {{{MET}} 00432},
  author = {Nakamura, T. and Noguchi, T. and Kimura, Y. and Hiroi, T. and Ahn, I. and Lee, J. I. and Sasaki, S.},
  year = {2013},
  month = sep,
  journal = {Meteoritics and Planetary Science Supplement},
  volume = {76},
  pages = {5122},
  urldate = {2025-08-28},
  adsurl = {https://ui.adsabs.harvard.edu/abs/2013M\&PSA..76.5122N}
}

@article{dermott2018,
  title = {The Common Origin of Family and Non-Family Asteroids},
  author = {Dermott, Stanley F. and Christou, Apostolos A. and Li, Dan and Kehoe, {\relax Thomas}. J. J. and Robinson, J. Malcolm},
  year = {2018},
  month = jul,
  journal = {Nature Astronomy},
  volume = {2},
  pages = {549--554},
  issn = {2397-3366},
  doi = {10.1038/s41550-018-0482-4},
  urldate = {2025-08-28},
  adsurl = {https://ui.adsabs.harvard.edu/abs/2018NatAs...2..549D}
}

@article{ferrone2023,
  title = {Identification of a 4.3 Billion Year Old Asteroid Family and Planetesimal Population in the {{Inner Main Belt}}},
  author = {Ferrone, S. and Delbo, M. and Avdellidou, C. and Melikyan, R. and Morbidelli, A. and Walsh, K. and Deienno, R.},
  year = {2023},
  month = aug,
  journal = {\aap},
  volume = {676},
  pages = {A5},
  publisher = {EDP Sciences},
  issn = {0004-6361, 1432-0746},
  doi = {10.1051/0004-6361/202245594},
  urldate = {2025-02-19},
  copyright = {{\copyright} The Authors 2023},
  langid = {english},
  adsurl = {https://www.aanda.org/articles/aa/abs/2023/08/aa45594-22/aa45594-22.html}
}

@article{neumann2012,
  title = {Differentiation and Core Formation in Accreting Planetesimals},
  author = {Neumann, W. and Breuer, D. and Spohn, T.},
  year = {2012},
  month = jul,
  journal = {\aap},
  volume = {543},
  pages = {A141},
  issn = {0004-6361},
  doi = {10.1051/0004-6361/201219157},
  urldate = {2025-08-28},
  adsurl = {https://ui.adsabs.harvard.edu/abs/2012A\&A...543A.141N}
}

@article{weiss2013,
  title = {Differentiated {{Planetesimals}} and the {{Parent Bodies}} of {{Chondrites}}},
  author = {Weiss, Benjamin P. and {Elkins-Tanton}, Linda T.},
  year = {2013},
  month = may,
  journal = {AREPS},
  volume = {41},
  pages = {529--560},
  issn = {0084-6597},
  doi = {10.1146/annurev-earth-040610-133520},
  urldate = {2025-06-17},
  adsurl = {https://ui.adsabs.harvard.edu/abs/2013AREPS..41..529W}
}

@phdthesis{bus1999,
  title = {Compositional Structure in the Asteroid Belt: {{Results}} of a Spectroscopic Survey},
  shorttitle = {Compositional Structure in the Asteroid Belt},
  author = {Bus, Schelte John},
  year = {1999},
  month = jan,
  pages = {311},
  urldate = {2025-08-28},
  school = {Massachusetts Institute of Technology}
}

@article{kallemeyn1986,
  title = {Compositions of Enstatite ({{EH3}}, {{EH4}},5 and {{EL6}}) Chondrites: {{Implications}} Regarding Their Formation},
  shorttitle = {Compositions of Enstatite ({{EH3}}, {{EH4}},5 and {{EL6}}) Chondrites},
  author = {Kallemeyn, Gregory W. and Wasson, John T.},
  year = {1986},
  month = oct,
  journal = {Geochimica et Cosmochimica Acta},
  volume = {50},
  pages = {2153--2164},
  issn = {0016-7037},
  doi = {10.1016/0016-7037(86)90070-0},
  urldate = {2025-08-28},
  adsurl = {https://ui.adsabs.harvard.edu/abs/1986GeCoA..50.2153K}
}

@article{zhang1995,
  title = {The Classification and Complex Thermal History of the Enstatite Chondrites},
  author = {Zhang, Yanhong and Benoit, Paul H. and Sears, Derek W. G.},
  year = {1995},
  month = may,
  journal = {\jgr},
  volume = {100},
  pages = {9417--9438},
  issn = {0148-0227},
  doi = {10.1029/95JE00502},
  urldate = {2025-08-28},
  adsurl = {https://ui.adsabs.harvard.edu/abs/1995JGR...100.9417Z}
}

@article{kruijer2017,
  title = {Age of {{Jupiter}} Inferred from the Distinct Genetics and Formation Times of Meteorites},
  author = {Kruijer, Thomas S. and Burkhardt, Christoph and Budde, Gerrit and Kleine, Thorsten},
  year = {2017},
  month = jun,
  journal = {Proceedings of the National Academy of Science},
  volume = {114},
  pages = {6712--6716},
  issn = {0027-8424},
  doi = {10.1073/pnas.1704461114},
  urldate = {2025-06-25},
  adsurl = {https://ui.adsabs.harvard.edu/abs/2017PNAS..114.6712K}
}

@phdthesis{tholen1984,
  title = {Asteroid {{Taxonomy}} from {{Cluster Analysis}} of {{Photometry}}.},
  author = {Tholen, David James},
  year = {1984},
  month = sep,
  urldate = {2025-08-28},
  school = {University of Arizona}
}

@article{bus2002a,
  title = {Phase {{II}} of the {{Small Main-Belt Asteroid Spectroscopic Survey}}. {{A Feature-Based Taxonomy}}},
  author = {Bus, Schelte J. and Binzel, Richard P.},
  year = {2002},
  month = jul,
  journal = {Icarus},
  volume = {158},
  pages = {146--177},
  issn = {0019-1035},
  doi = {10.1006/icar.2002.6856},
  urldate = {2025-08-28},
  adsurl = {https://ui.adsabs.harvard.edu/abs/2002Icar..158..146B}
}

@article{fujiya2019,
  title = {Migration of {{D-type}} Asteroids from the Outer {{Solar System}} Inferred from Carbonate in Meteorites},
  author = {Fujiya, W. and Hoppe, P. and Ushikubo, T. and Fukuda, K. and Lindgren, P. and Lee, M. R. and Koike, M. and Shirai, K. and Sano, Y.},
  year = {2019},
  month = jul,
  journal = {Nature Astronomy},
  volume = {3},
  pages = {910--915},
  issn = {2397-3366},
  doi = {10.1038/s41550-019-0801-4},
  urldate = {2025-08-28},
  adsurl = {https://ui.adsabs.harvard.edu/abs/2019NatAs...3..910F}
}

@article{usui2013,
  title = {Albedo {{Properties}} of {{Main Belt Asteroids Based}} on the {{All-Sky Survey}} of the {{Infrared Astronomical Satellite AKARI}}},
  author = {Usui, Fumihiko and Kasuga, Toshihiro and Hasegawa, Sunao and Ishiguro, Masateru and Kuroda, Daisuke and M{\"u}ller, Thomas G. and Ootsubo, Takafumi and Matsuhara, Hideo},
  year = {2013},
  month = jan,
  journal = {\apj},
  volume = {762},
  pages = {56},
  issn = {0004-637X},
  doi = {10.1088/0004-637X/762/1/56},
  urldate = {2025-08-28},
  adsurl = {https://ui.adsabs.harvard.edu/abs/2013ApJ...762...56U}
}

@article{durech2011,
  title = {Combining Asteroid Models Derived by Lightcurve Inversion with Asteroidal Occultation Silhouettes},
  author = {{\v D}urech, Josef and Kaasalainen, Mikko and Herald, David and Dunham, David and Timerson, Brad and Hanu{\v s}, Josef and Frappa, Eric and Talbot, John and Hayamizu, Tsutomu and Warner, Brian D. and Pilcher, Frederick and Gal{\'a}d, Adri{\'a}n},
  year = {2011},
  month = aug,
  journal = {Icarus},
  volume = {214},
  pages = {652--670},
  issn = {0019-1035},
  doi = {10.1016/j.icarus.2011.03.016},
  urldate = {2025-08-29},
  adsurl = {https://ui.adsabs.harvard.edu/abs/2011Icar..214..652D}
}

@article{hanus2017,
  title = {Volumes and Bulk Densities of Forty Asteroids from {{ADAM}} Shape Modeling},
  author = {Hanu{\v s}, J. and Viikinkoski, M. and Marchis, F. and {\v D}urech, J. and Kaasalainen, M. and Delbo', M. and Herald, D. and Frappa, E. and Hayamizu, T. and Kerr, S. and Preston, S. and Timerson, B. and Dunham, D. and Talbot, J.},
  year = {2017},
  month = may,
  journal = {\aap},
  volume = {601},
  pages = {A114},
  issn = {0004-6361},
  doi = {10.1051/0004-6361/201629956},
  urldate = {2025-08-29},
  adsurl = {https://ui.adsabs.harvard.edu/abs/2017A\&A...601A.114H}
}

@article{mainzer2019,
  title = {{{NEOWISE Diameters}} and {{Albedos V2}}.0},
  author = {Mainzer, Amy K. and Bauer, James M. and Cutri, Roc M. and Grav, Tommy and Kramer, Emily A. and Masiero, Joseph R. and Sonnett, Sarah and Wright, Edward L.},
  year = {2019},
  month = jan,
  journal = {NASA PDS},
  volume = {251},
  doi = {10.26033/18S3-2Z54},
  urldate = {2025-08-29},
  adsurl = {https://ui.adsabs.harvard.edu/abs/2019PDSS..251.....M}
}

@article{masiero2011,
  title = {Main {{Belt Asteroids}} with {{WISE}}/{{NEOWISE}}. {{I}}. {{Preliminary Albedos}} and {{Diameters}}},
  author = {Masiero, Joseph R. and Mainzer, A. K. and Grav, T. and Bauer, J. M. and Cutri, R. M. and Dailey, J. and Eisenhardt, P. R. M. and McMillan, R. S. and Spahr, T. B. and Skrutskie, M. F. and Tholen, D. and Walker, R. G. and Wright, E. L. and DeBaun, E. and Elsbury, D. and Gautier, IV, T. and Gomillion, S. and Wilkins, A.},
  year = {2011},
  month = nov,
  journal = {\apj},
  volume = {741},
  pages = {68},
  issn = {0004-637X},
  doi = {10.1088/0004-637X/741/2/68},
  urldate = {2025-08-29},
  adsurl = {https://ui.adsabs.harvard.edu/abs/2011ApJ...741...68M}
}

@article{ryan2010,
  title = {Rectified {{Asteroid Albedos}} and {{Diameters}} from {{IRAS}} and {{MSX Photometry Catalogs}}},
  author = {Ryan, Erin Lee and Woodward, Charles E.},
  year = {2010},
  month = oct,
  journal = {\aj},
  volume = {140},
  pages = {933--943},
  issn = {0004-6256},
  doi = {10.1088/0004-6256/140/4/933},
  urldate = {2025-08-29},
  adsurl = {https://ui.adsabs.harvard.edu/abs/2010AJ....140..933R}
}

@article{tedesco2002,
  title = {The {{Midcourse Space Experiment Infrared Minor Planet Survey}}},
  author = {Tedesco, Edward F. and Egan, Michael P. and Price, Stephan D.},
  year = {2002},
  month = jul,
  journal = {\aj},
  volume = {124},
  pages = {583--591},
  issn = {0004-6256},
  doi = {10.1086/340960},
  urldate = {2025-08-29},
  adsurl = {https://ui.adsabs.harvard.edu/abs/2002AJ....124..583T}
}

@article{usui2011a,
  title = {Asteroid {{Catalog Using Akari}}: {{AKARI}}/{{IRC Mid-Infrared Asteroid Survey}}},
  shorttitle = {Asteroid {{Catalog Using Akari}}},
  author = {Usui, Fumihiko and Kuroda, Daisuke and M{\"u}ller, Thomas G. and Hasegawa, Sunao and Ishiguro, Masateru and Ootsubo, Takafumi and Ishihara, Daisuke and Kataza, Hirokazu and Takita, Satoshi and Oyabu, Shinki and Ueno, Munetaka and Matsuhara, Hideo and Onaka, Takashi},
  year = {2011},
  month = oct,
  journal = {\pasj},
  volume = {63},
  pages = {1117--1138},
  issn = {0004-6264},
  doi = {10.1093/pasj/63.5.1117},
  urldate = {2025-08-29},
  adsurl = {https://ui.adsabs.harvard.edu/abs/2011PASJ...63.1117U}
}

@inproceedings{knezevic2012,
  title = {Asteroids {{Dynamic Site-AstDyS}}},
  booktitle = {{{IAU Joint Discussion}}},
  author = {Knezevic, Zoran and Milani, Andrea},
  year = {2012},
  month = aug,
  pages = {P18},
  urldate = {2025-08-29}
}

@article{budde2016,
  title = {Molybdenum Isotopic Evidence for the Origin of Chondrules and a Distinct Genetic Heritage of Carbonaceous and Non-Carbonaceous Meteorites},
  author = {Budde, Gerrit and Burkhardt, Christoph and Brennecka, Gregory A. and {Fischer-G{\"o}dde}, Mario and Kruijer, Thomas S. and Kleine, Thorsten},
  year = {2016},
  month = nov,
  journal = {E\&PSL},
  volume = {454},
  pages = {293--303},
  issn = {0012-821X},
  doi = {10.1016/j.epsl.2016.09.020},
  urldate = {2025-09-01},
  adsurl = {https://ui.adsabs.harvard.edu/abs/2016E\&PSL.454..293B}
}

@article{galinier2024,
  title = {Discovery of the First Olivine-Dominated {{A-type}} Asteroid Family},
  author = {Galinier, M. and Delbo, M. and Avdellidou, C. and Galluccio, L.},
  year = {2024},
  month = mar,
  journal = {Astronomy \& Astrophysics},
  volume = {683},
  pages = {L3},
  publisher = {EDP Sciences},
  issn = {0004-6361, 1432-0746},
  doi = {10.1051/0004-6361/202349057},
  urldate = {2025-02-20},
  copyright = {{\copyright} The Authors 2024},
  langid = {english},
  adsurl = {https://www.aanda.org/articles/aa/abs/2024/03/aa49057-23/aa49057-23.html}
}

@article{oszkiewicz2015,
  title = {Differentiation Signatures in the {{Flora}} Region},
  author = {Oszkiewicz, Dagmara and Kankiewicz, Pawe{\l} and W{\l}odarczyk, Ireneusz and Kryszczy{\'n}ska, Agnieszka},
  year = {2015},
  month = dec,
  journal = {Astronomy and Astrophysics},
  volume = {584},
  pages = {A18},
  issn = {0004-6361},
  doi = {10.1051/0004-6361/201526219},
  urldate = {2025-09-10},
  adsurl = {https://ui.adsabs.harvard.edu/abs/2015A\&A...584A..18O}
}

@article{gradie1982,
  title = {Compositional {{Structure}} of the {{Asteroid Belt}}},
  author = {Gradie, J. and Tedesco, E.},
  year = {1982},
  month = jun,
  journal = {Science},
  volume = {216},
  pages = {1405--1407},
  issn = {0036-8075},
  doi = {10.1126/science.216.4553.1405},
  urldate = {2025-09-10},
  adsurl = {https://ui.adsabs.harvard.edu/abs/1982Sci...216.1405G}
}

@article{ciocco2025,
  title = {A Collisional History of the {{L}} Chondrite Parent Bodies},
  author = {Ciocco, Marine and Roskosz, Mathieu and Doisneau, B{\'e}atrice and Deloule, Etienne and Fiquet, Guillaume and Delbo, Marco and Gounelle, Matthieu},
  year = {2025},
  month = sep,
  journal = {Nature Astronomy},
  issn = {2397-3366},
  doi = {10.1038/s41550-025-02615-6},
  urldate = {2025-09-10},
  langid = {english},
  adsurl = {https://www.nature.com/articles/s41550-025-02615-6}
}

@article{galinier2023,
  title = {Gaia Search for Early-Formed Andesitic Asteroidal Crusts},
  author = {Galinier, M. and Delbo, M. and Avdellidou, C. and Galluccio, L. and Marrocchi, Y.},
  year = {2023},
  month = mar,
  journal = {Astronomy and Astrophysics},
  volume = {671},
  pages = {A40},
  issn = {0004-6361},
  doi = {10.1051/0004-6361/202245311},
  urldate = {2025-04-24},
  adsurl = {https://ui.adsabs.harvard.edu/abs/2023A\&A...671A..40G}
}

@article{nakamura2023,
  title = {Formation and Evolution of Carbonaceous Asteroid {{Ryugu}}: {{Direct}} Evidence from Returned Samples},
  shorttitle = {Formation and Evolution of Carbonaceous Asteroid {{Ryugu}}},
  author = {Nakamura, T. and Matsumoto, M. and Amano, K. and Enokido, Y. and Zolensky, M. E. and Mikouchi, T. and Genda, H. and Tanaka, S. and Zolotov, M. Y. and Kurosawa, K. and Wakita, S. and Hyodo, R. and Nagano, H. and Nakashima, D. and Takahashi, Y. and Fujioka, Y. and Kikuiri, M. and Kagawa, E. and Matsuoka, M. and Brearley, A. J. and Tsuchiyama, A. and Uesugi, M. and Matsuno, J. and Kimura, Y. and Sato, M. and Milliken, R. E. and Tatsumi, E. and Sugita, S. and Hiroi, T. and Kitazato, K. and Brownlee, D. and Joswiak, D. J. and Takahashi, M. and Ninomiya, K. and Takahashi, T. and Osawa, T. and Terada, K. and Brenker, F. E. and Tkalcec, B. J. and Vincze, L. and Brunetto, R. and {Al{\'e}on-Toppani}, A. and Chan, Q. H. S. and Roskosz, M. and Viennet, J. -C. and Beck, P. and Alp, E. E. and Michikami, T. and Nagaashi, Y. and Tsuji, T. and Ino, Y. and Martinez, J. and Han, J. and Dolocan, A. and Bodnar, R. J. and Tanaka, M. and Yoshida, H. and Sugiyama, K. and King, A. J. and Fukushi, K. and Suga, H. and Yamashita, S. and Kawai, T. and Inoue, K. and Nakato, A. and Noguchi, T. and Vilas, F. and Hendrix, A. R. and {Jaramillo-Correa}, C. and Domingue, D. L. and Dominguez, G. and Gainsforth, Z. and Engrand, C. and Duprat, J. and Russell, S. S. and Bonato, E. and Ma, C. and Kawamoto, T. and Wada, T. and Watanabe, S. and Endo, R. and Enju, S. and Riu, L. and Rubino, S. and Tack, P. and Takeshita, S. and Takeichi, Y. and Takeuchi, A. and Takigawa, A. and Takir, D. and Tanigaki, T. and Taniguchi, A. and Tsukamoto, K. and Yagi, T. and Yamada, S. and Yamamoto, K. and Yamashita, Y. and Yasutake, M. and Uesugi, K. and Umegaki, I. and Chiu, I. and Ishizaki, T. and Okumura, S. and Palomba, E. and Pilorget, C. and Potin, S. M. and Alasli, A. and Anada, S. and Araki, Y. and Sakatani, N. and Schultz, C. and Sekizawa, O. and Sitzman, S. D. and Sugiura, K. and Sun, M. and Dartois, E. and De Pauw, E. and Dionnet, Z. and Djouadi, Z. and Falkenberg, G. and Fujita, R. and Fukuma, T. and Gearba, I. R. and Hagiya, K. and Hu, M. Y. and Kato, T. and Kawamura, T. and Kimura, M. and Kubo, M. K. and Langenhorst, F. and Lantz, C. and Lavina, B. and Lindner, M. and Zhao, J. and Vekemans, B. and Baklouti, D. and Bazi, B. and Borondics, F. and Nagasawa, S. and Nishiyama, G. and Nitta, K. and Mathurin, J. and Matsumoto, T. and Mitsukawa, I. and Miura, H. and Miyake, A. and Miyake, Y. and Yurimoto, H. and Okazaki, R. and Yabuta, H. and Naraoka, H. and Sakamoto, K. and Tachibana, S. and Connolly, H. C. and Lauretta, D. S. and Yoshitake, M. and Yoshikawa, M. and Yoshikawa, K. and Yoshihara, K. and Yokota, Y. and Yogata, K. and Yano, H. and Yamamoto, Y. and Yamamoto, D. and Yamada, M. and Yamada, T. and Yada, T. and Wada, K. and Usui, T. and Tsukizaki, R. and Terui, F. and Takeuchi, H. and Takei, Y. and Iwamae, A. and Soejima, H. and Shirai, K. and Shimaki, Y. and Senshu, H. and Sawada, H. and Saiki, T. and Ozaki, M. and Ono, G. and Okada, T. and Ogawa, N. and Ogawa, K. and Noguchi, R. and Noda, H. and Nishimura, M. and Namiki, N. and Nakazawa, S. and Morota, T. and Miyazaki, A. and Miura, A. and Mimasu, Y. and Matsumoto, K. and Kumagai, K. and Kouyama, T. and Kikuchi, S. and Kawahara, K. and Kameda, S.},
  year = {2023},
  month = mar,
  journal = {Science},
  volume = {379},
  pages = {abn8671},
  issn = {0036-8075},
  doi = {10.1126/science.abn8671},
  urldate = {2025-06-25},
  adsurl = {https://ui.adsabs.harvard.edu/abs/2023Sci...379.8671N}
}

@article{gradie1980,
  title = {The Composition of the {{Trojan}} Asteroids},
  author = {Gradie, J. and Veverka, J.},
  year = {1980},
  month = feb,
  journal = {Nature},
  volume = {283},
  pages = {840--842},
  issn = {0028-0836},
  doi = {10.1038/283840a0},
  urldate = {2025-06-25},
  adsurl = {https://ui.adsabs.harvard.edu/abs/1980Natur.283..840G}
}

@techreport{nationalresearchcouncil2007,
  title = {Exploring {{Organic Environments}} in the {{Solar System}}},
  author = {{National Research Council}},
  year = {2007},
  month = jan,
  journal = {National Research Council. 2007. Exploring Organic Environments in the Solar System. Washington},
  doi = {10.17226/11860},
  urldate = {2025-09-12}
}

@incollection{glavin2015,
title = {Chapter 3 - The Origin and Evolution of Organic Matter in Carbonaceous Chondrites and Links to Their Parent Bodies},
editor = {Neyda Abreu},
booktitle = {Primitive Meteorites and Asteroids},
publisher = {Elsevier},
pages = {205-271},
year = {2018},
isbn = {978-0-12-813325-5},
doi = {https://doi.org/10.1016/B978-0-12-813325-5.00003-3},
url = {https://www.sciencedirect.com/science/article/pii/B9780128133255000033},
author = {Daniel P. Glavin and Conel M.O'D. Alexander and José C. Aponte and Jason P. Dworkin and Jamie E. Elsila and Hikaru Yabuta},
}

@article{burkhardt2016,
  title = {A Nucleosynthetic Origin for the {{Earth}}'s Anomalous {{142Nd}} Composition},
  author = {Burkhardt, C. and Borg, L. E. and Brennecka, G. A. and Shollenberger, Q. R. and Dauphas, N. and Kleine, T.},
  year = {2016},
  month = sep,
  journal = {Nature},
  volume = {537},
  pages = {394--398},
  issn = {0028-0836},
  doi = {10.1038/nature18956},
  urldate = {2025-09-12},
  adsurl = {https://ui.adsabs.harvard.edu/abs/2016Natur.537..394B}
}

@inproceedings{novakovic2019,
  title = {Asteroid Families Portal},
  booktitle = {{{EPSC-DPS Joint Meeting}} 2019},
  author = {Novakovic, Bojan and Radovic, Viktor},
  year = 2019,
  month = sep,
  volume = {2019},
  pages = {EPSC-DPS2019-1671},
  urldate = {2025-11-10},
  langid = {english}
}

@misc{delbo2025,
  title = {Gaia and {{IRTF}} Abundance of A-Type Main Belt Asteroids},
  author = {Delbo, Marco and Avdellidou, Chrysa and Galinier, Marjorie and Bhat, Ullas and Dyer, Thomas and Bolin, Bryce T. and Galluccio, Laurent},
  year = 2025,
  month = nov,
  publisher = {arXiv},
  doi = {10.48550/arXiv.2511.00902},
  urldate = {2025-11-10},
  langid = {english}
}

\clearpage
\appendix
\section{TABLES AND FIGURES}
\onecolumn
\begin{footnotesize}

\begin{landscape}
\begin{longtable}{|l|cccccccc|cccccccc|}
\caption{List of asteroid families studied. The albedos are retrieved from MP3C. The first- and second-best BDM class, along with their corresponding $\chi^2$ values, of the family average reflectance spectra in the NES15 and AFP25 catalogues are provided. The confidence value for a margin of error of 10\% of the calculated average reflectance spectra based on the proportion of the Gaia DR3 samples compared to the full family, as described in Section \ref{sec:z-score} is provided ranging from 0 to 1.}
\label{tab:family-list} \\
\hline
\multirow{3}{*}{Family} & \multicolumn{8}{c|}{NES15} & \multicolumn{8}{c|}{AFP25} \\*
 & \multicolumn{3}{c}{$N$} & \multicolumn{2}{c}{$p_\text{V}$} & \multirow{2}{*}{Class} & $\chi^2_1,\chi^2_2$ & \multirow{2}{*}{Conf.} & \multicolumn{3}{c}{$N$} & \multicolumn{2}{c}{$p_\text{V}$} & \multirow{2}{*}{Class} & $\chi^2_1,\chi^2_2$ & \multirow{2}{*}{Conf.} \\*
 & All & Gaia & SDSS & All & Gaia &  & $\times 10^{-4}$ &  & All & Gaia & SDSS & All & Gaia &  & $\times 10^{-4}$ &  \endfirsthead 
\hline
10 Hygiea & 4854 & 494 & 321 & 0.07±0.03 & 0.08±0.03 & Cb, Ch & 1.76, 2.61 & 1.0 & 10716 & 485 & 315 & 0.07±0.03 & 0.08±0.03 & Cb, Ch & 2.50, 3.41 & 1.0 \\
22 Kalliope & 300 & 36 & 26 & 0.19±0.06 & 0.2±0.05 & X, Xc & 1.57, 4.17 & 0.8 & - & - & - & - & - & -, - & -, - & - \\
24 Themis & 4782 & 1055 & 805 & 0.07±0.02 & 0.08±0.02 & Ch, Cb & 0.28, 1.05 & 1.0 & 12288 & 1130 & 860 & 0.07±0.02 & 0.08±0.02 & Ch, Cb & 0.35, 1.01 & 1.0 \\
31 Euphrosyne & 2035 & 354 & 121 & 0.06±0.02 & 0.06±0.03 & Cb, Cgh & 2.13, 3.40 & 1.0 & 12260 & 410 & 134 & 0.06±0.03 & 0.06±0.03 & Cb, Cgh & 2.52, 2.83 & 1.0 \\
81 Terpsichore & 138 & 8 & 12 & 0.06±0.02 & 0.06±0.02 & Xc, X & 2.75, 3.03 & 0.44 & 405 & 14 & 12 & 0.06±0.03 & 0.09±0.06 & Xc, X & 0.96, 1.24 & 0.55 \\
86 Semele & - & - & - & - & - & -, - & -, - & - & 468 & 1 & 0 & 0.06±0.04 & 0.05±- & Cgh, Cb & 3.41, 5.69 & 0.16 \\
87 Sylvia & 255 & 25 & 7 & 0.05±0.02 & 0.06±0.03 & X, Xc & 0.55, 3.37 & 0.71 & 1249 & 18 & 6 & 0.06±0.03 & 0.07±0.03 & X, T & 4.35, 9.36 & 0.61 \\
96 Aegle & 99 & 32 & 27 & 0.07±0.02 & 0.07±0.02 & D, T & 2.00, 6.99 & 0.83 & 171 & 33 & 29 & 0.07±0.02 & 0.07±0.02 & D, T & 1.92, 7.57 & 0.8 \\
128 Nemesis & 1302 & 51 & 45 & 0.08±0.03 & 0.11±0.06 & Cgh, Cb & 2.96, 4.05 & 0.85 & - & - & - & - & - & -, - & -, - & - \\
137 Meliboea & 444 & 119 & 73 & 0.06±0.03 & 0.06±0.03 & Cgh, Cb & 3.92, 5.07 & 0.99 & - & - & - & - & - & -, - & -, - & - \\
144 Vibilia & 135 & 30 & 22 & 0.08±0.05 & 0.09±0.08 & X, Xc & 4.48, 6.91 & 0.79 & - & - & - & - & - & -, - & -, - & - \\
145 Adeona & 2236 & 404 & 319 & 0.07±0.04 & 0.07±0.03 & Ch, Cb & 3.32, 3.46 & 1.0 & 6279 & 407 & 314 & 0.07±0.04 & 0.07±0.04 & Cgh, Cb & 3.58, 3.63 & 1.0 \\
283 Emma & 442 & 50 & 30 & 0.05±0.02 & 0.05±0.03 & X, Xc & 1.36, 2.68 & 0.87 & 1093 & 47 & 27 & 0.05±0.02 & 0.05±0.03 & X, Xc & 1.43, 2.60 & 0.84 \\
293 Brasilia & 579 & 64 & 42 & 0.17±0.06 & 0.2±0.08 & X, Xc & 2.00, 3.83 & 0.91 & - & - & - & - & - & -, - & -, - & - \\
322 Phaeo & 146 & 27 & 13 & 0.06±0.02 & 0.07±0.02 & T, X & 8.79, 11.24 & 0.75 & 335 & 10 & 6 & 0.08±0.06 & 0.11±0.04 & T, X & 6.14, 7.78 & 0.48 \\
363 Padua & 1087 & 135 & 109 & 0.07±0.03 & 0.08±0.04 & X, Xc & 1.67, 5.08 & 0.99 & 1469 & 96 & 76 & 0.07±0.02 & 0.07±0.03 & X, Xc & 1.89, 5.45 & 0.96 \\
369 Aeria & 272 & 15 & 8 & 0.16±0.08 & 0.18±0.11 & X, Xc & 1.63, 4.64 & 0.57 & 807 & 14 & 9 & 0.17±0.07 & 0.15±0.05 & X, Xc & 1.65, 4.77 & 0.55 \\
375 Ursula & 1466 & 285 & 211 & 0.07±0.04 & 0.08±0.05 & X, Xc & 2.47, 3.54 & 1.0 & 6082 & 296 & 221 & 0.07±0.04 & 0.08±0.05 & X, Xc & 2.75, 3.73 & 1.0 \\
396 Aeolia & 296 & 8 & 13 & 0.1±0.03 & 0.1±0.03 & X, Xc & 2.70, 4.45 & 0.43 & 1683 & 15 & 26 & 0.1±0.06 & 0.13±0.09 & X, Xc & 2.83, 5.31 & 0.56 \\
410 Chloris & 424 & 78 & 88 & 0.1±0.07 & 0.09±0.04 & X, Xc & 5.44, 8.15 & 0.95 & 753 & 56 & 79 & 0.1±0.07 & 0.09±0.04 & X, Xc & 5.48, 8.11 & 0.88 \\
490 Veritas & 1294 & 135 & 154 & 0.07±0.02 & 0.07±0.02 & Cgh, X & 5.09, 5.13 & 0.99 & 5789 & 140 & 158 & 0.07±0.03 & 0.07±0.03 & Cgh, X & 5.03, 5.11 & 0.98 \\
569 Misa & 702 & 53 & 54 & 0.06±0.03 & 0.06±0.02 & X, Xc & 3.70, 4.48 & 0.87 & 1959 & 33 & 12 & 0.06±0.03 & 0.06±0.02 & Cgh, X & 4.04, 4.63 & 0.75 \\
589 Croatia & 161 & 12 & 3 & 0.06±0.02 & 0.07±0.03 & X, Xc & 1.83, 5.44 & 0.53 & - & - & - & - & - & -, - & -, - & - \\
618 Elfriede & 63 & 2 & 0 & 0.05±0.02 & 0.07±0.03 & Cgh, Cb & 9.49, 12.19 & 0.23 & 527 & 2 & 1 & 0.07±0.03 & 0.05±0.0 & T, X & 3.86, 8.22 & 0.22 \\
627 Charis & 808 & 20 & 20 & 0.12±0.09 & 0.2±0.12 & T, D & 2.16, 6.79 & 0.63 & 2586 & 17 & 19 & 0.12±0.09 & 0.21±0.11 & T, D & 2.36, 7.01 & 0.59 \\
656 Beagle & 148 & 24 & 15 & 0.07±0.02 & 0.08±0.02 & Ch, Cb & 0.24, 1.40 & 0.72 & 830 & 76 & 58 & 0.07±0.02 & 0.08±0.02 & Ch, Cb & 0.24, 1.22 & 0.93 \\
668 Dora & 1259 & 248 & 154 & 0.05±0.02 & 0.05±0.01 & Cb, Ch & 3.80, 4.07 & 1.0 & 3872 & 244 & 149 & 0.06±0.02 & 0.05±0.01 & Cb, Ch & 3.87, 4.19 & 1.0 \\
709 Fringilla & 134 & 32 & 24 & 0.05±0.03 & 0.06±0.05 & X, Xc & 1.85, 4.90 & 0.81 & 200 & 4 & 1 & 0.06±0.04 & 0.07±0.05 & X, Xc & 3.22, 6.88 & 0.31 \\
727 Nipponia & - & - & - & - & - & -, - & -, - & - & 355 & 9 & 3 & 0.17±0.08 & 0.18±0.08 & Xc, X & 0.58, 0.87 & 0.46 \\
778 Theobalda & 376 & 43 & 31 & 0.06±0.02 & 0.07±0.02 & Ch, Cb & 2.42, 4.28 & 0.84 & 3185 & 44 & 31 & 0.06±0.02 & 0.07±0.02 & Ch, Cb & 2.22, 4.05 & 0.82 \\
780 Armenia & 40 & 2 & 1 & 0.06±0.02 & 0.06±0.02 & X, T & 24.22, 27.58 & 0.23 & 422 & 3 & 1 & 0.06±0.03 & 0.08±0.04 & T, X & 9.36, 11.24 & 0.27 \\
816 Juliana & 76 & 14 & 3 & 0.07±0.06 & 0.14±0.07 & X, Xc & 0.63, 2.91 & 0.59 & - & - & - & - & - & -, - & -, - & - \\
845 Naema & 301 & 30 & 32 & 0.06±0.02 & 0.07±0.01 & Cgh, X & 4.31, 6.24 & 0.75 & 1227 & 34 & 33 & 0.06±0.02 & 0.07±0.01 & Cgh, Cb & 4.18, 6.38 & 0.76 \\
909 Ulla & 26 & 5 & 3 & 0.06±0.03 & 0.06±0.02 & X, Xc & 6.72, 12.69 & 0.38 & 97 & 5 & 3 & 0.06±0.03 & 0.06±0.02 & X, Xc & 6.22, 11.97 & 0.35 \\
926 Imhilde & 43 & 3 & 4 & 0.05±0.01 & 0.06±0.02 & X, Xc & 7.30, 10.69 & 0.28 & - & - & - & - & - & -, - & -, - & - \\
1128 Astrid & 489 & 13 & 7 & 0.05±0.01 & 0.06±0.01 & X, Xc & 3.66, 5.60 & 0.54 & 1349 & 13 & 7 & 0.05±0.02 & 0.06±0.01 & X, Xc & 3.71, 5.68 & 0.53 \\
1189 Terentia & 79 & 3 & 0 & 0.06±0.03 & 0.07±0.03 & X, Xc & 6.65, 9.15 & 0.28 & 421 & 3 & 0 & 0.06±0.03 & 0.07±0.03 & X, Xc & 6.68, 9.36 & 0.27 \\
1222 Tina & 96 & 9 & 1 & 0.13±0.04 & 0.14±0.05 & X, Xc & 1.14, 4.04 & 0.47 & 720 & 9 & 1 & 0.12±0.05 & 0.14±0.05 & X, Xc & 1.30, 4.38 & 0.45 \\
1303 Luthera & 163 & 24 & 27 & 0.05±0.02 & 0.05±0.01 & T, D & 6.55, 9.49 & 0.71 & 878 & 31 & 27 & 0.06±0.03 & 0.06±0.03 & T, D & 6.22, 10.09 & 0.74 \\
1332 Marconia & 34 & 4 & 0 & 0.04±0.01 & 0.05±0.01 & X, Xc & 4.23, 4.83 & 0.33 & - & - & - & - & - & -, - & -, - & - \\
1484 Postrema & 108 & 19 & 14 & 0.06±0.05 & 0.07±0.05 & D, T & 2.70, 4.27 & 0.66 & 128 & 10 & 1 & 0.05±0.02 & 0.06±0.05 & T, Xe & 1.31, 10.87 & 0.49 \\
1521 Seinajoki & - & - & - & - & - & -, - & -, - & - & 2824 & 64 & 42 & 0.16±0.06 & 0.2±0.08 & X, Xc & 2.06, 3.88 & 0.89 \\
1668 Hanna & 280 & 14 & 3 & 0.06±0.04 & 0.07±0.06 & Cgh, X & 5.63, 6.82 & 0.56 & 1007 & 12 & 3 & 0.06±0.04 & 0.07±0.06 & Cgh, X & 5.29, 7.38 & 0.51 \\
1726 Hoffmeister & 1819 & 67 & 56 & 0.05±0.02 & 0.05±0.01 & Cgh, Cb & 5.25, 5.89 & 0.9 & 5319 & 67 & 51 & 0.05±0.02 & 0.05±0.02 & T, X & 8.62, 12.31 & 0.9 \\
2262 Mitidika & 653 & 182 & 84 & 0.06±0.02 & 0.07±0.02 & X, Xc & 3.71, 4.76 & 1.0 & - & - & - & - & - & -, - & -, - & - \\
2782 Leonidas & 135 & 29 & 21 & 0.08±0.05 & 0.08±0.07 & X, Xc & 4.84, 7.36 & 0.78 & 237 & 21 & 5 & 0.07±0.05 & 0.08±0.08 & X, Xc & 5.81, 7.86 & 0.66 \\
3438 Inarradas & 38 & 15 & 15 & 0.08±0.02 & 0.08±0.02 & Cgh, Ch & 3.22, 4.13 & 0.68 & - & - & - & - & - & -, - & -, - & - \\
3556 Lixiaohua & 756 & 66 & 40 & 0.04±0.01 & 0.04±0.01 & X, Xc & 1.35, 3.53 & 0.91 & 4913 & 78 & 45 & 0.05±0.03 & 0.05±0.03 & X, Xc & 2.51, 4.65 & 0.93 \\
3811 Karma & 124 & 13 & 6 & 0.07±0.06 & 0.09±0.08 & Xc, Xk & 1.19, 1.56 & 0.55 & 517 & 15 & 6 & 0.07±0.05 & 0.11±0.1 & Xc, X & 1.41, 1.59 & 0.57 \\
3815 Konig & 354 & 15 & 10 & 0.05±0.03 & 0.04±0.01 & Cgh, X & 8.14, 8.75 & 0.57 & 1994 & 15 & 10 & 0.05±0.03 & 0.04±0.01 & X, Xc & 4.35, 8.56 & 0.56 \\
4203 Brucato & 342 & 110 & 41 & 0.06±0.02 & 0.06±0.02 & Cgh, Xc & 3.36, 4.69 & 0.99 & 212 & 19 & 3 & 0.07±0.03 & 0.07±0.03 & Cb, Ch & 3.36, 4.88 & 0.64 \\
5438 Lorre & 2 & 1 & 0 & 0.05±0.0 & 0.05±- & T, X & 3.57, 9.56 & 0.22 & 93 & 1 & 0 & 0.06±0.03 & 0.05±- & T, X & 3.57, 9.56 & 0.16 \\
5567 Durisen & 27 & 5 & 14 & 0.04±0.01 & 0.04±0.01 & T, X & 4.53, 4.74 & 0.38 & - & - & - & - & - & -, - & -, - & - \\
5614 Yakovlev & 67 & 9 & 9 & 0.05±0.01 & 0.05±0.01 & X, Xc & 2.24, 3.69 & 0.48 & 273 & 13 & 11 & 0.06±0.03 & 0.05±0.02 & X, Xc & 1.88, 2.55 & 0.54 \\
15454 1998YB3 & 38 & 14 & 12 & 0.05±0.01 & 0.05±0.01 & X, Cgh & 6.34, 6.52 & 0.65 & - & - & - & - & - & -, - & -, - & - \\
18405 1993FY12 & 104 & 10 & 9 & 0.17±0.05 & 0.15±0.05 & X, Xc & 2.63, 5.87 & 0.49 & 449 & 10 & 9 & 0.16±0.06 & 0.15±0.05 & X, Xc & 2.61, 5.72 & 0.48 \\
53546 2000BY6 & 58 & 9 & 2 & 0.11±- & 0.11±- & D, T & 9.82, 13.38 & 0.49 & 236 & 10 & 2 & 0.13±0.11 & 0.11±- & D, T & 10.26, 12.25 & 0.48 \\
106302 2000UJ87 & - & - & - & - & - & -, - & -, - & - & - & - & - & - & - & -, - & -, - & - \\
\hline
\end{longtable}
\end{landscape}
\end{footnotesize}

\begin{figure*}
	\centering
	\includegraphics[width=\linewidth]{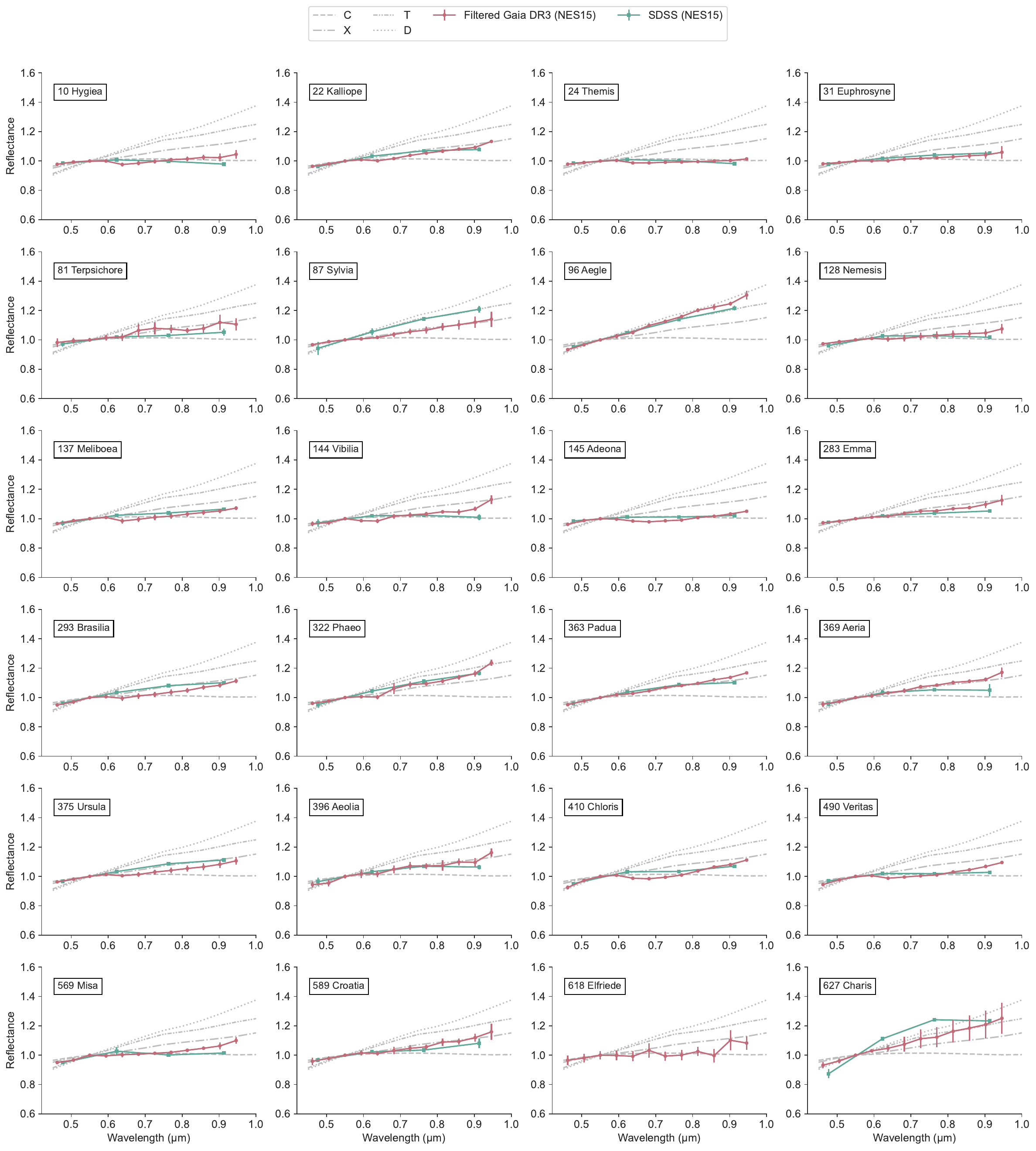}
	\caption{The average reflectance Gaia DR3 spectra (red) of the asteroid families in NES15. The spectra are normalised at 0.55~\SI{}{\micro\meter}. The average SDSS spectra (teal) are also shown for comparison, when available.}
	\label{fig:all-spectra-nes}
\end{figure*}

\begin{figure*}
	\centering
	\includegraphics[width=\linewidth]{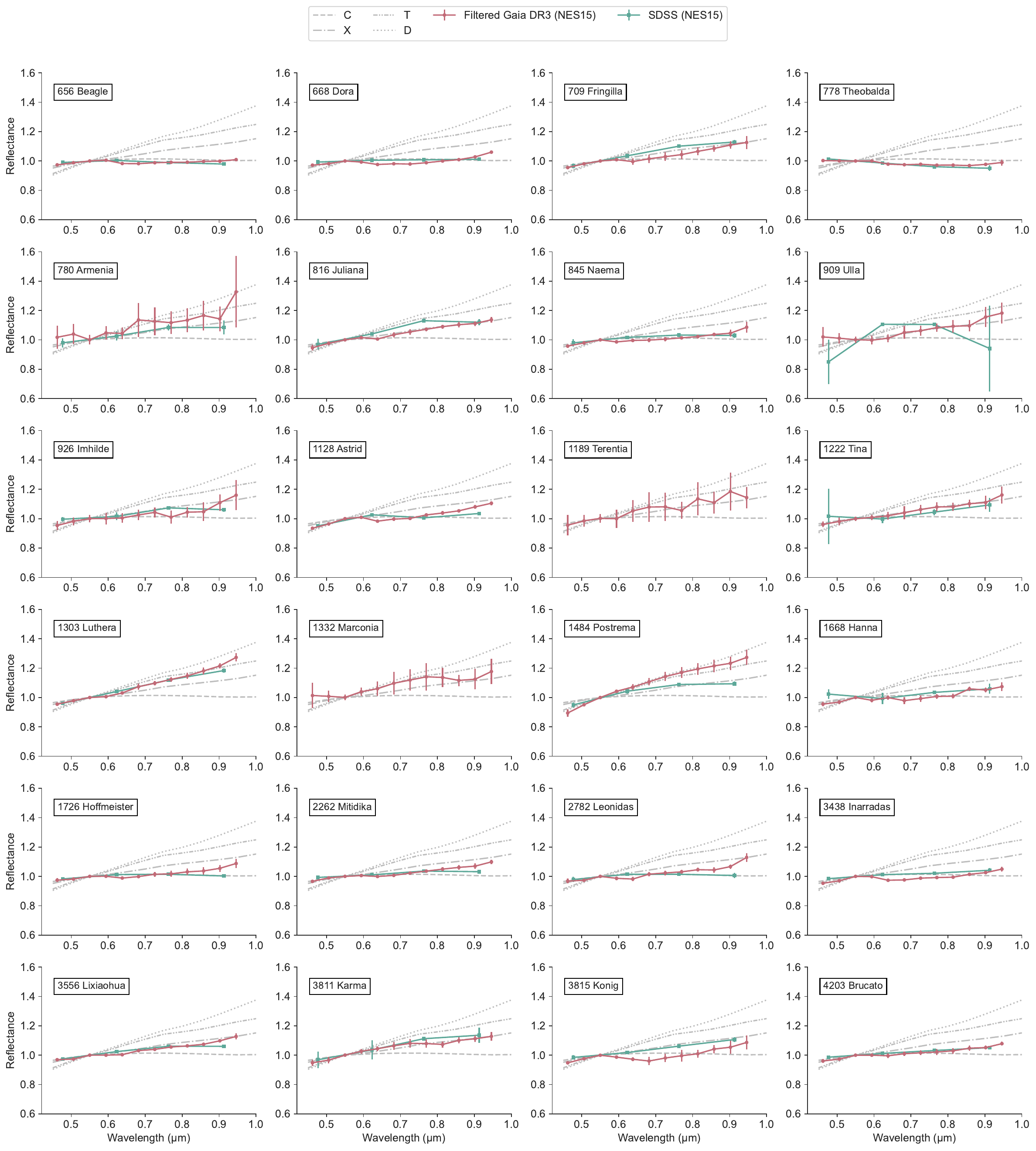}
	\contcaption{The average reflectance Gaia DR3 spectra (red) of the asteroid families  in NES15. The spectra are normalised at 0.55~\SI{}{\micro\meter}. The average SDSS spectra (teal) are also shown for comparison, when available.}
\end{figure*}

\begin{figure*}
	\centering
	\includegraphics[width=\linewidth]{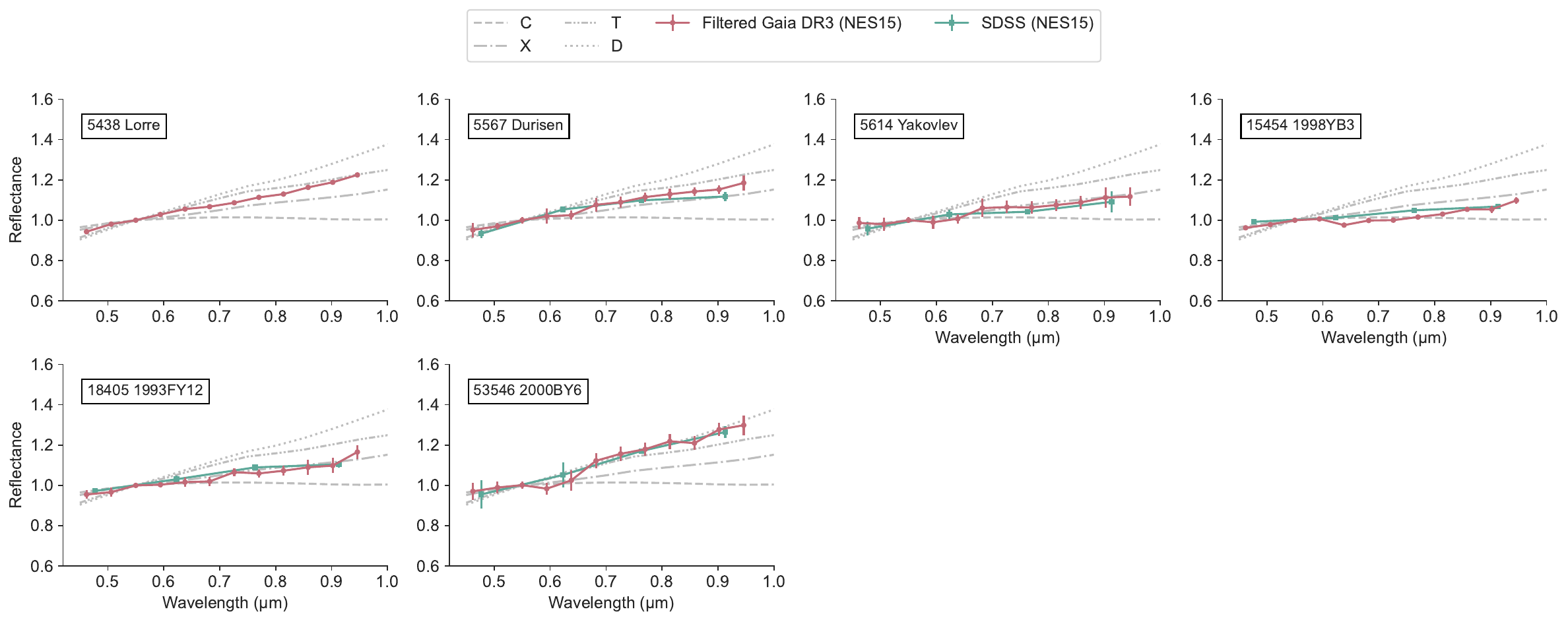}
	\contcaption{The average reflectance Gaia DR3 spectra (red) of the asteroid families in NES15. The spectra are normalised at 0.55~\SI{}{\micro\meter}. The average SDSS spectra (teal) are also shown for comparison, when available.}
\end{figure*}

\begin{figure*}
	\centering
	\includegraphics[width=\linewidth]{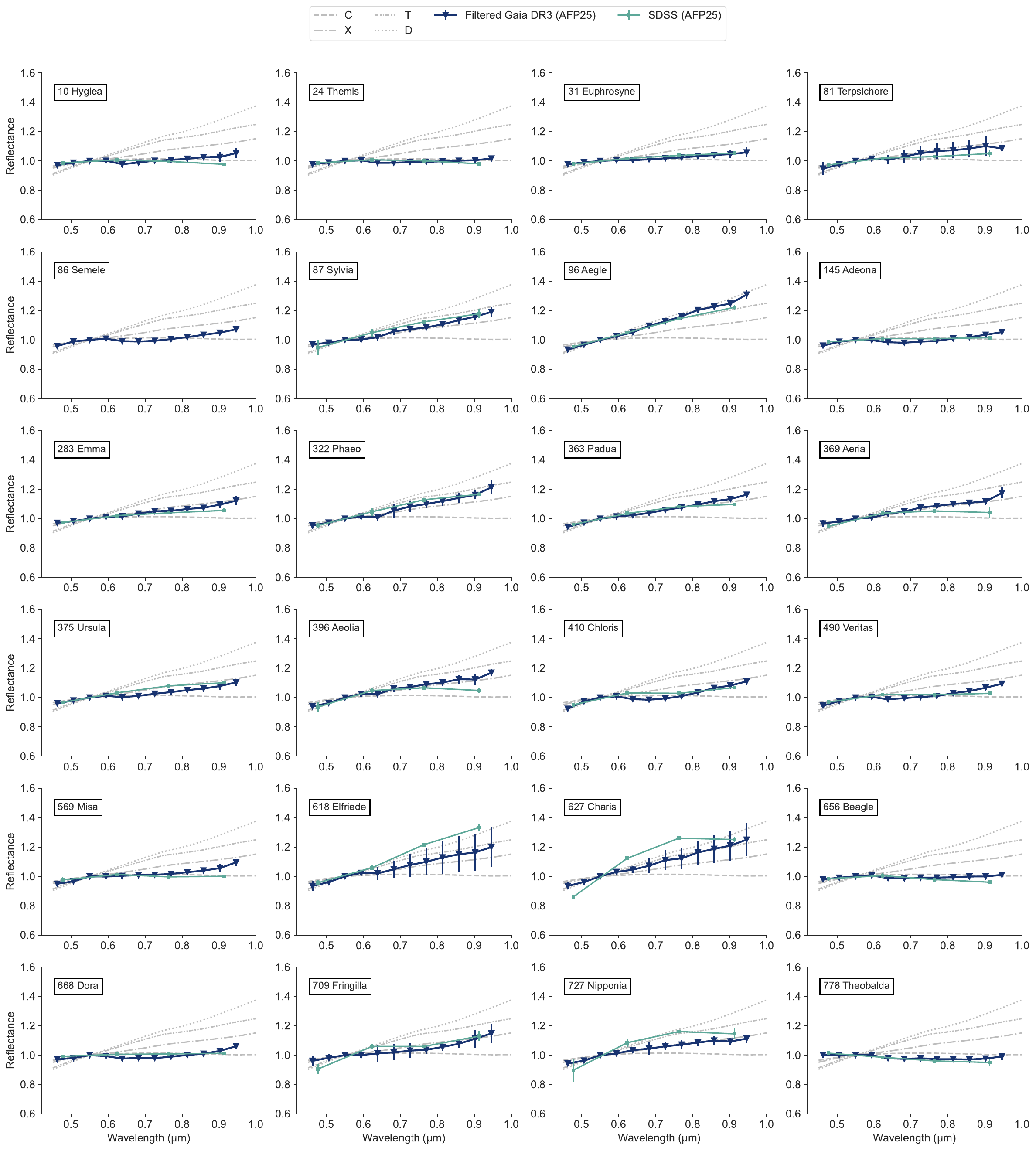}
	\caption{The average reflectance Gaia DR3 spectra (blue) of the asteroid families in AFP25. The spectra are normalised at 0.55~\SI{}{\micro\meter}. The average SDSS spectra (teal) are also shown for comparison, when available.}
	\label{fig:all-spectra-nov}
\end{figure*}

\begin{figure*}
	\centering
	\includegraphics[width=\linewidth]{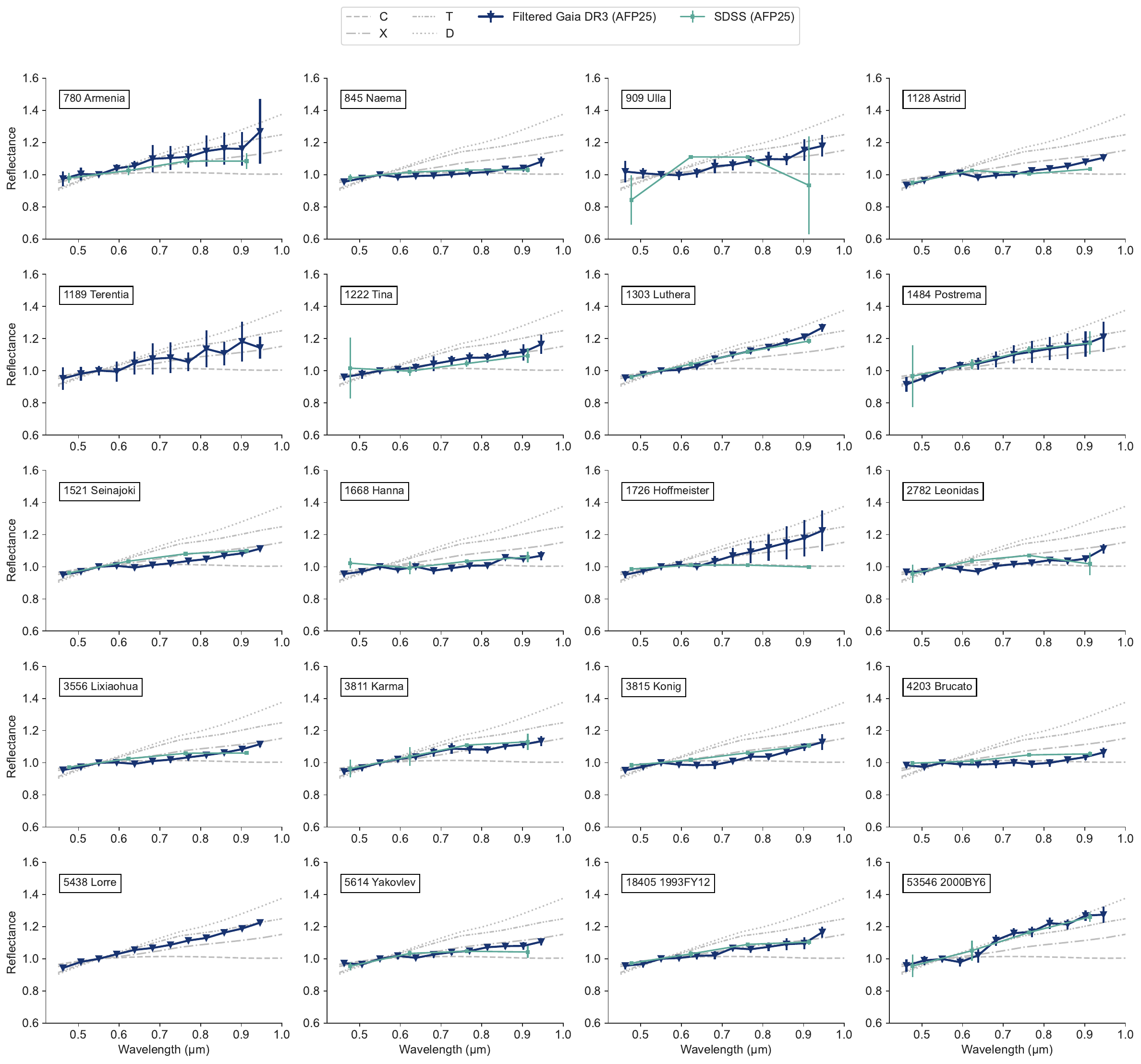}
	\contcaption{The average reflectance Gaia DR3 spectra (blue) of the asteroid families in AFP25. The spectra are normalised at 0.55~\SI{}{\micro\meter}. The average SDSS spectra (teal) are also shown for comparison, when available.}
\end{figure*}

\begin{figure*}
	\centering
	\includegraphics[width=\linewidth]{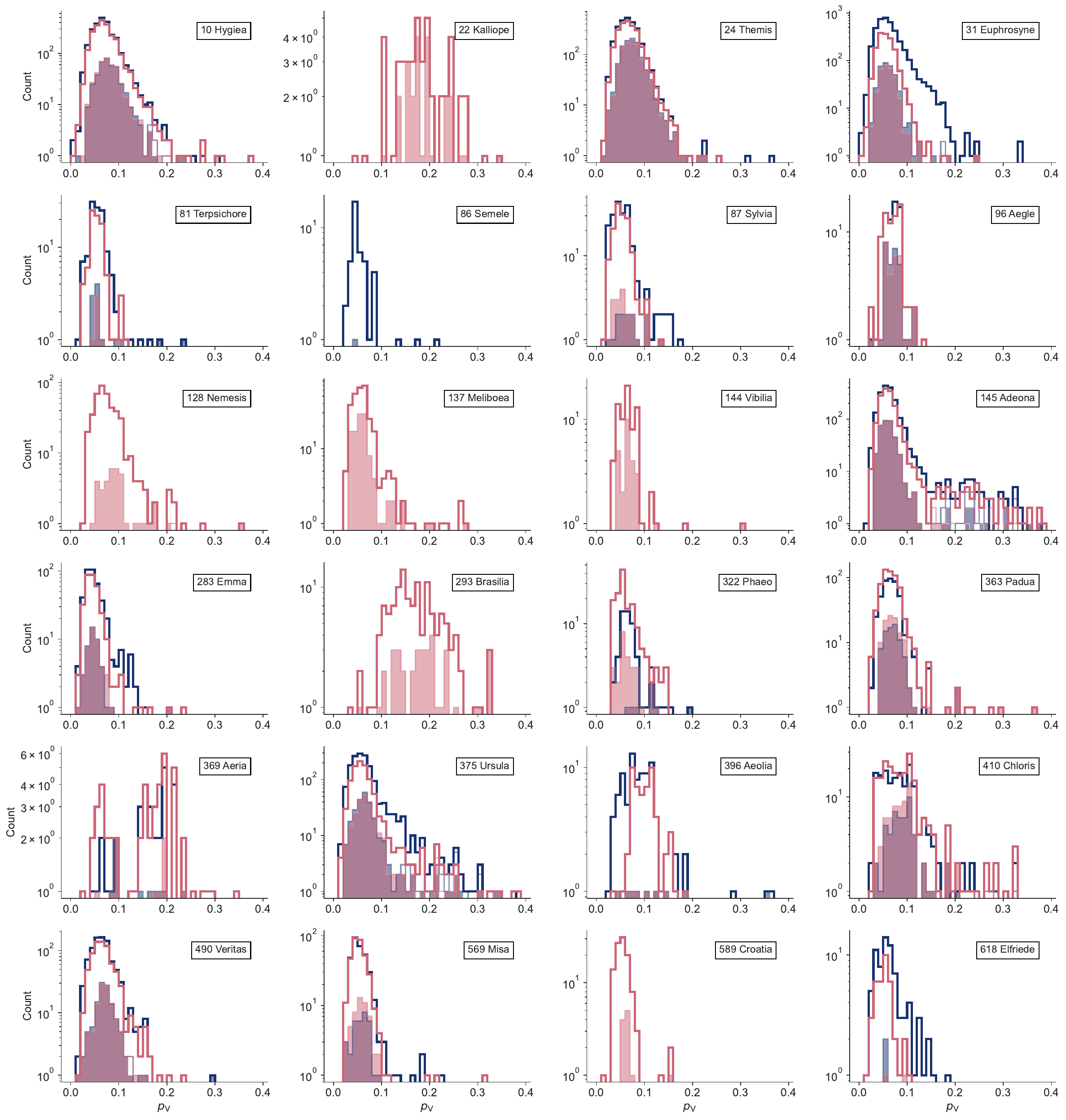}
	\caption{The albedo distribution of C- and X-complex asteroid families. The thick and thin red lines represent the albedo distribution of all members and members with Gaia DR3 spectra respectively in the NES15 catalogue, with the shaded red region represents the filtered family list from NES15 as described in Section \ref{sec:interlopers}. The blue lines and shaded regions represent the same for the AFP25 catalogue.}
	\label{fig:all-albedo}
\end{figure*}

\begin{figure*}
	\centering
	\includegraphics[width=\linewidth]{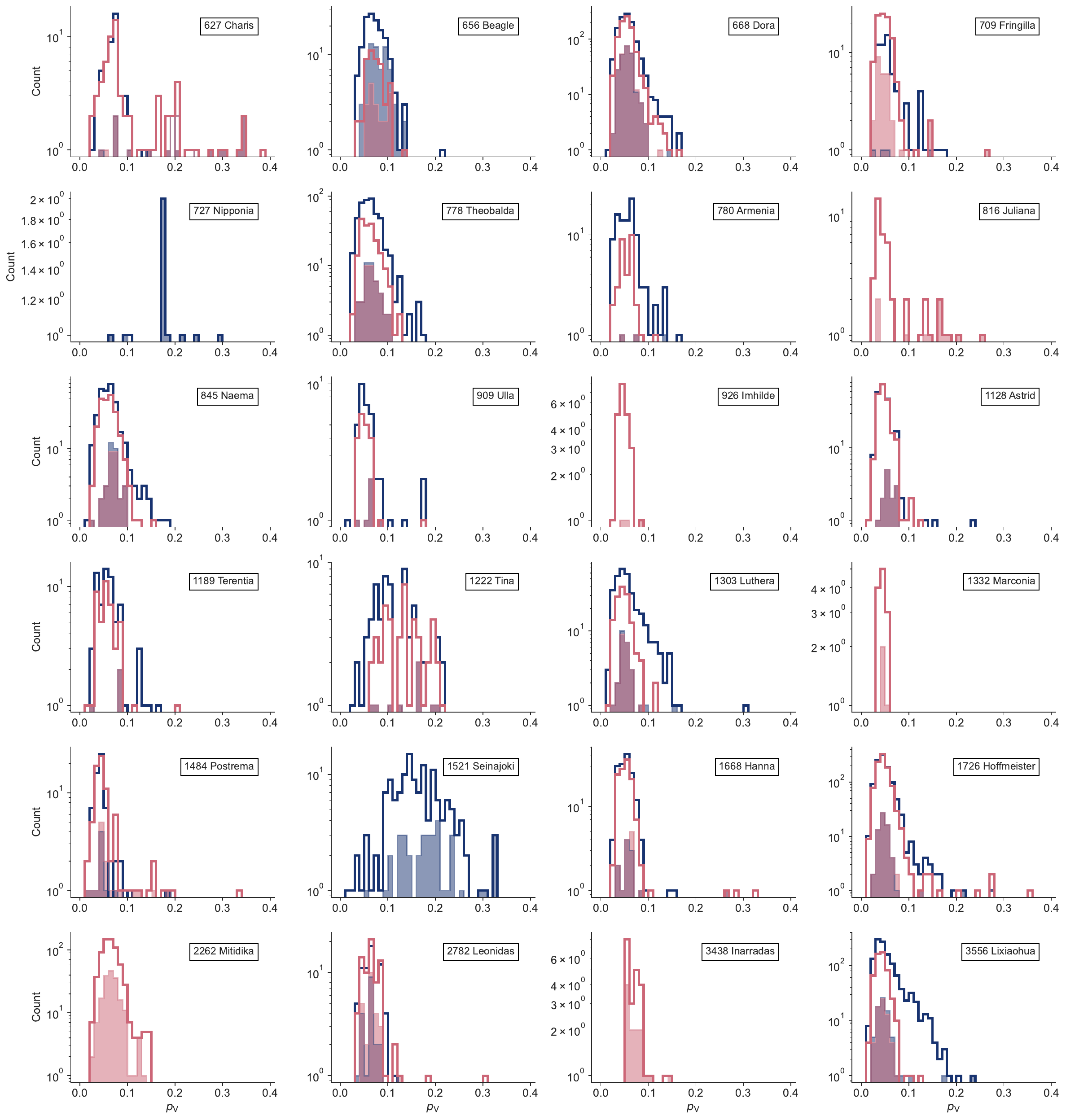}
	\contcaption{The albedo distribution of C- and X-complex asteroid families. The thick and thin red lines represent the albedo distribution of all members and members with Gaia DR3 spectra respectively in the NES15 catalogue, with the shaded red region represents the filtered family list from NES15 as described in Section \ref{sec:interlopers}. The blue lines and shaded regions represent the same for the AFP25 catalogue.}
\end{figure*}

\begin{figure*}
	\centering
	\includegraphics[width=\linewidth]{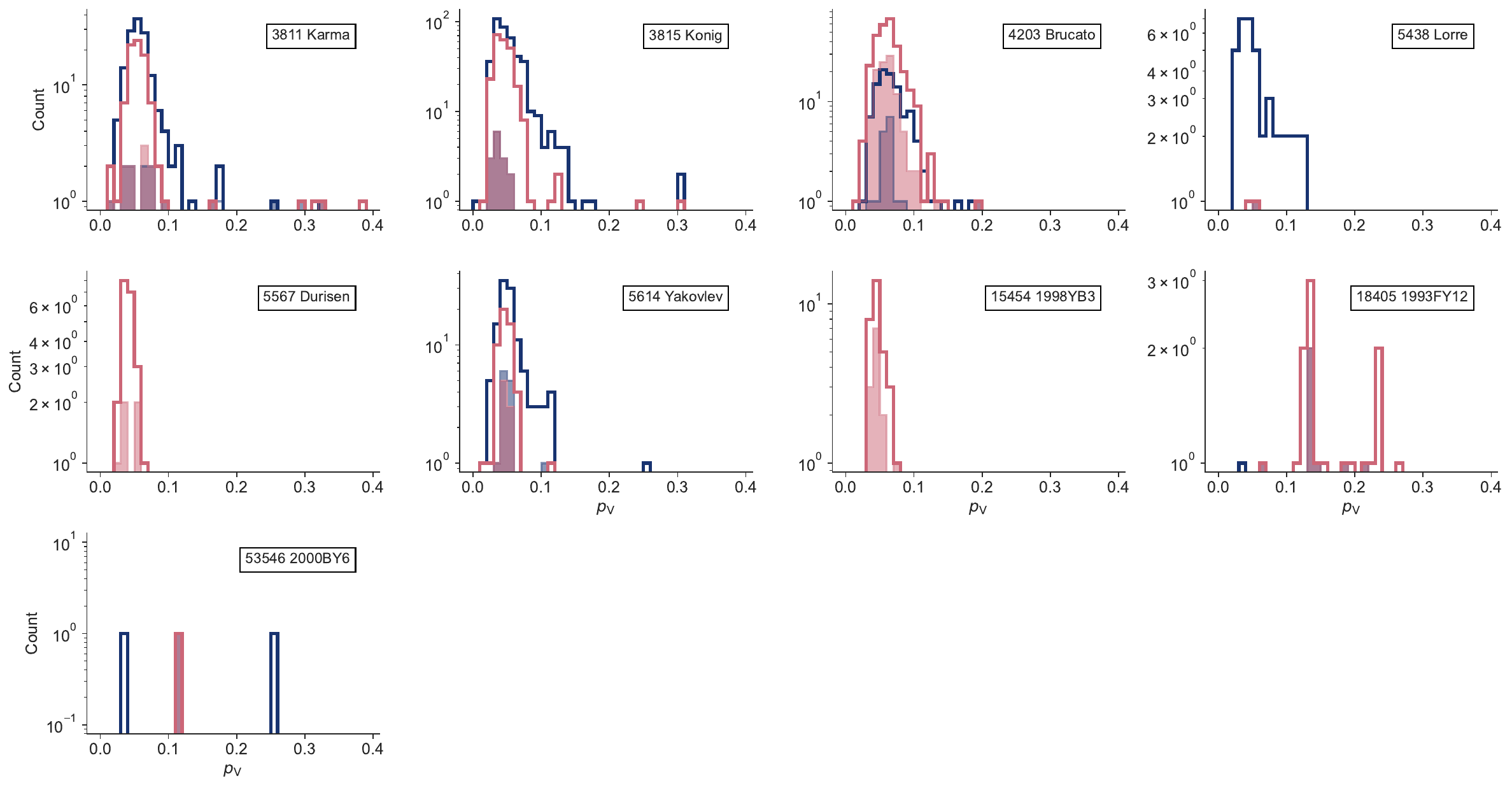}
	\contcaption{The albedo distribution of C- and X-complex asteroid families. The thick and thin red lines represent the albedo distribution of all members and members with Gaia DR3 spectra respectively in the NES15 catalogue, with the shaded red region represents the filtered family list from NES15 as described in Section \ref{sec:interlopers}. The blue lines and shaded regions represent the same for the AFP25 catalogue.}
\end{figure*}

\begin{figure*}
	\centering
	\includegraphics[width=\textwidth]{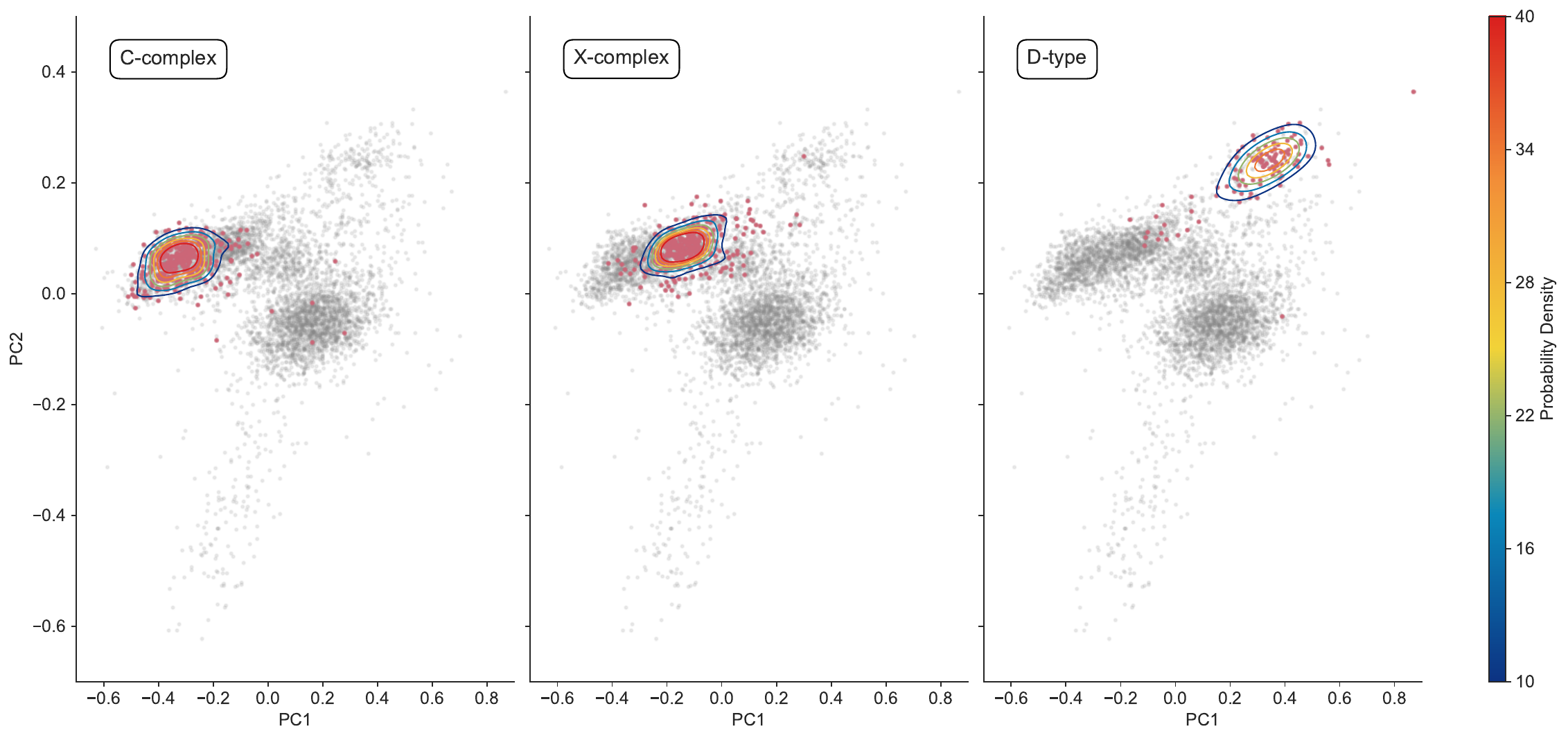}
	\caption{PC1-PC2 distribution of the Gaia DR3. The grey dots represent the distribution of asteroids with $\mathrm{SNR}>75$. The red dots represent the spectroscopic C-complex (left), X-complex (middle) and D-type (right) asteroids. The KDEs for the distribution of the C-complex (left), X-complex (middle) and D-type (right) asteroids are shown as contours.}
	\label{fig:gaia-pca}
\end{figure*}



\bsp	
\label{lastpage}
\end{document}